\documentclass[aps, prd, twocolumn,superscriptaddress]{revtex4-2}

\usepackage{amsmath}
\usepackage{float}
\usepackage{subfig}
\usepackage{amsfonts}
\usepackage{amssymb}
\usepackage{graphicx}
\usepackage{siunitx}
\usepackage{chngpage}
\usepackage[left]{lineno}%
\RequirePackage{graphicx}

\usepackage[margin=0.8in]{geometry}
\usepackage{dcolumn}
\usepackage{bm}
\usepackage{lipsum}
\usepackage{mathptmx}
\usepackage[usenames, dvipsnames]{color}
\usepackage[pdftex]{hyperref}
\providecommand{\U}[1]{\protect\rule{.1in}{.1in}}
\usepackage{enumerate}
\begin{document}

\widetext

\title{Identification and reconstruction of low-energy electrons in the ProtoDUNE-SP detector}
%

\newcommand{\Abilene}{Abilene Christian University, Abilene, TX 79601, USA}
\newcommand{\Albanysuny}{University of Albany, SUNY, Albany, NY 12222, USA}
\newcommand{\Amsterdam}{University of Amsterdam, NL-1098 XG Amsterdam, The Netherlands}
\newcommand{\Antalya}{Antalya Bilim University, 07190 D{\"o}{\c{s}}emealt{\i}/Antalya, Turkey}
\newcommand{\Antananarivo}{University of Antananarivo, Antananarivo 101, Madagascar}
\newcommand{\AntonioNarino}{Universidad Antonio Nari{\~n}o, Bogot{\'a}, Colombia}
\newcommand{\Argonne}{Argonne National Laboratory, Argonne, IL 60439, USA}
\newcommand{\Arizona}{University of Arizona, Tucson, AZ 85721, USA}
\newcommand{\Asuncion}{Universidad Nacional de Asunci{\'o}n, San Lorenzo, Paraguay}
\newcommand{\Athens}{University of Athens, Zografou GR 157 84, Greece}
\newcommand{\Atlantico}{Universidad del Atl{\'a}ntico, Barranquilla, Atl{\'a}ntico, Colombia}
\newcommand{\Augustana}{Augustana University, Sioux Falls, SD 57197, USA}
\newcommand{\Basel}{University of Basel, CH-4056 Basel, Switzerland}
\newcommand{\Bern}{University of Bern, CH-3012 Bern, Switzerland}
\newcommand{\Beykent}{Beykent University, Istanbul, Turkey}
\newcommand{\Birmingham}{University of Birmingham, Birmingham B15 2TT, United Kingdom}
\newcommand{\BolognaUniversity}{Universit{\`a} del Bologna, 40127 Bologna, Italy}
\newcommand{\Boston}{Boston University, Boston, MA 02215, USA}
\newcommand{\Bristol}{University of Bristol, Bristol BS8 1TL, United Kingdom}
\newcommand{\Brookhaven}{Brookhaven National Laboratory, Upton, NY 11973, USA}
\newcommand{\Bucharest}{University of Bucharest, Bucharest, Romania}
\newcommand{\CalBerkeley}{University of California Berkeley, Berkeley, CA 94720, USA}
\newcommand{\CalDavis}{University of California Davis, Davis, CA 95616, USA}
\newcommand{\CalIrvine}{University of California Irvine, Irvine, CA 92697, USA}
\newcommand{\CalLosangeles}{University of California Los Angeles, Los Angeles, CA 90095, USA}
\newcommand{\CalRiverside}{University of California Riverside, Riverside CA 92521, USA}
\newcommand{\CalSantabarbara}{University of California Santa Barbara, Santa Barbara, California 93106 USA}
\newcommand{\Caltech}{California Institute of Technology, Pasadena, CA 91125, USA}
\newcommand{\Cambridge}{University of Cambridge, Cambridge CB3 0HE, United Kingdom}
\newcommand{\Campinas}{Universidade Estadual de Campinas, Campinas - SP, 13083-970, Brazil}
\newcommand{\CataniaUniversitadi}{Universit{\`a} di Catania, 2 - 95131 Catania, Italy}
\newcommand{\Catolica}{Universidad Cat{\'o}lica del Norte, Antofagasta, Chile}
\newcommand{\CBPF}{Centro Brasileiro de Pesquisas F\'isicas, Rio de Janeiro, RJ 22290-180, Brazil}
\newcommand{\CEASaclay}{IRFU, CEA, Universit{\'e} Paris-Saclay, F-91191 Gif-sur-Yvette, France}
\newcommand{\CERN}{CERN, The European Organization for Nuclear Research, 1211 Meyrin, Switzerland}
\newcommand{\Charles}{Institute of Particle and Nuclear Physics of the Faculty of Mathematics and Physics of the Charles University, 180 00 Prague 8, Czech Republic }
\newcommand{\Chicago}{University of Chicago, Chicago, IL 60637, USA}
\newcommand{\ChungAng}{Chung-Ang University, Seoul 06974, South Korea}
\newcommand{\CIEMAT}{CIEMAT, Centro de Investigaciones Energ{\'e}ticas, Medioambientales y Tecnol{\'o}gicas, E-28040 Madrid, Spain}
\newcommand{\Cincinnati}{University of Cincinnati, Cincinnati, OH 45221, USA}
\newcommand{\Cinvestav}{Centro de Investigaci{\'o}n y de Estudios Avanzados del Instituto Polit{\'e}cnico Nacional (Cinvestav), Mexico City, Mexico}
\newcommand{\Colima}{Universidad de Colima, Colima, Mexico}
\newcommand{\ColoradoBoulder}{University of Colorado Boulder, Boulder, CO 80309, USA}
\newcommand{\ColoradoState}{Colorado State University, Fort Collins, CO 80523, USA}
\newcommand{\Columbia}{Columbia University, New York, NY 10027, USA}
\newcommand{\conida}{Comisi{\'o}n Nacional de Investigaci{\'o}n y Desarrollo Aeroespacial, Lima, Peru}
\newcommand{\Cti}{Centro de Tecnologia da Informacao Renato Archer, Amarais - Campinas, SP - CEP 13069-901}
\newcommand{\CUSB}{Central University of South Bihar, Gaya, 824236, India }
\newcommand{\CzechAcademyofSciences}{Institute of Physics, Czech Academy of Sciences, 182 00 Prague 8, Czech Republic}
\newcommand{\CzechTechnical}{Czech Technical University, 115 19 Prague 1, Czech Republic}
\newcommand{\DakotaState}{Dakota State University, Madison, SD 57042, USA}
\newcommand{\Dallas}{University of Dallas, Irving, TX 75062-4736, USA}
\newcommand{\DannecyleVieux}{Laboratoire d{\textquoteright}Annecy de Physique des Particules, Universit{\'e} Grenoble Alpes, Universit{\'e} Savoie Mont Blanc, CNRS, LAPP-IN2P3, 74000 Annecy, France}
\newcommand{\Daresbury}{Daresbury Laboratory, Cheshire WA4 4AD, United Kingdom}
\newcommand{\Drexel}{Drexel University, Philadelphia, PA 19104, USA}
\newcommand{\Duke}{Duke University, Durham, NC 27708, USA}
\newcommand{\Durham}{Durham University, Durham DH1 3LE, United Kingdom}
\newcommand{\Edinburgh}{University of Edinburgh, Edinburgh EH8 9YL, United Kingdom}
\newcommand{\EIA}{Universidad EIA, Envigado, Antioquia, Colombia}
\newcommand{\ETH}{ETH Zurich, Zurich, Switzerland}
\newcommand{\Eotvos}{Eotvos Lor{\'a}nd University, 1053 Budapest, Hungary}
\newcommand{\FCULport}{Faculdade de Ci{\^e}ncias da Universidade de Lisboa - FCUL, 1749-016 Lisboa, Portugal}
\newcommand{\FederaldeAlfenas}{Universidade Federal de Alfenas, Po{\c{c}}os de Caldas - MG, 37715-400, Brazil}
\newcommand{\FederaldeGoias}{Universidade Federal de Goias, Goiania, GO 74690-900, Brazil}
\newcommand{\FederaldoABC}{Universidade Federal do ABC, Santo Andr{\'e} - SP, 09210-580, Brazil}
\newcommand{\FederaldoRio}{Universidade Federal do Rio de Janeiro,  Rio de Janeiro - RJ, 21941-901, Brazil}
\newcommand{\Fermi}{Fermi National Accelerator Laboratory, Batavia, IL 60510, USA}
\newcommand{\Ferrarauniv}{University of Ferrara, Ferrara, Italy}
\newcommand{\Florida}{University of Florida, Gainesville, FL 32611-8440, USA}
\newcommand{\Floridastate}{Florida State University, Tallahassee, FL 32306, USA}
\newcommand{\Fluminense}{Fluminense Federal University, 9 Icara{\'\i} Niter{\'o}i - RJ, 24220-900, Brazil }
\newcommand{\Genova}{Universit{\`a} degli Studi di Genova, Genova, Italy}
\newcommand{\Georgian}{Georgian Technical University, Tbilisi, Georgia}
\newcommand{\Granada}{University of Granada {\&} CAFPE, 18002 Granada, Spain}
\newcommand{\GranSasso}{Gran Sasso Science Institute, L'Aquila, Italy}
\newcommand{\GranSassoLab}{Laboratori Nazionali del Gran Sasso, L'Aquila AQ, Italy}
\newcommand{\Grenoble}{University Grenoble Alpes, CNRS, Grenoble INP, LPSC-IN2P3, 38000 Grenoble, France}
\newcommand{\Guanajuato}{Universidad de Guanajuato, Guanajuato, C.P. 37000, Mexico}
\newcommand{\Harish}{Harish-Chandra Research Institute, Jhunsi, Allahabad 211 019, India}
\newcommand{\Harvard}{Harvard University, Cambridge, MA 02138, USA}
\newcommand{\Hawaii}{University of Hawaii, Honolulu, HI 96822, USA}
\newcommand{\Houston}{University of Houston, Houston, TX 77204, USA}
\newcommand{\Hyderabad}{University of  Hyderabad, Gachibowli, Hyderabad - 500 046, India}
\newcommand{\Idaho}{Idaho State University, Pocatello, ID 83209, USA}
\newcommand{\IFAE}{Institut de F{\'\i}sica d{\textquoteright}Altes Energies (IFAE){\textemdash}Barcelona Institute of Science and Technology (BIST), Barcelona, Spain}
\newcommand{\IFIC}{Instituto de F{\'\i}sica Corpuscular, CSIC and Universitat de Val{\`e}ncia, 46980 Paterna, Valencia, Spain}
\newcommand{\IGFAE}{Instituto Galego de F{\'\i}sica de Altas Enerx{\'\i}as, University of Santiago de Compostela, Santiago de Compostela, 15782, Spain}
\newcommand{\Illinoisinstitute}{Illinois Institute of Technology, Chicago, IL 60616, USA}
\newcommand{\Imperial}{Imperial College of Science Technology and Medicine, London SW7 2BZ, United Kingdom}
\newcommand{\IndGuwahati}{Indian Institute of Technology Guwahati, Guwahati, 781 039, India}
\newcommand{\IndHyderabad}{Indian Institute of Technology Hyderabad, Hyderabad, 502285, India}
\newcommand{\Indiana}{Indiana University, Bloomington, IN 47405, USA}
\newcommand{\INFNBologna}{Istituto Nazionale di Fisica Nucleare Sezione di Bologna, 40127 Bologna BO, Italy}
\newcommand{\INFNCatania}{Istituto Nazionale di Fisica Nucleare Sezione di Catania, I-95123 Catania, Italy}
\newcommand{\INFNFerrara}{Istituto Nazionale di Fisica Nucleare Sezione di Ferrara, I-44122 Ferrara, Italy}
\newcommand{\INFNGenova}{Istituto Nazionale di Fisica Nucleare Sezione di Genova, 16146 Genova GE, Italy}
\newcommand{\INFNLecce}{Istituto Nazionale di Fisica Nucleare Sezione di Lecce, 73100 - Lecce, Italy}
\newcommand{\INFNMilanBicocca}{Istituto Nazionale di Fisica Nucleare Sezione di Milano Bicocca, 3 - I-20126 Milano, Italy}
\newcommand{\INFNMilano}{Istituto Nazionale di Fisica Nucleare Sezione di Milano, 20133 Milano, Italy}
\newcommand{\INFNNapoli}{Istituto Nazionale di Fisica Nucleare Sezione di Napoli, I-80126 Napoli, Italy}
\newcommand{\INFNPadova}{Istituto Nazionale di Fisica Nucleare Sezione di Padova, 35131 Padova, Italy}
\newcommand{\INFNPavia}{Istituto Nazionale di Fisica Nucleare Sezione di Pavia,  I-27100 Pavia, Italy}
\newcommand{\INFNPisa}{Istituto Nazionale di Fisica Nucleare Laboratori Nazionali di Pisa, Pisa PI, Italy}
\newcommand{\INFNRoma}{Istituto Nazionale di Fisica Nucleare Sezione di Roma, 00185 Roma RM, Italy}
\newcommand{\INFNSud}{Istituto Nazionale di Fisica Nucleare Laboratori Nazionali del Sud, 95123 Catania, Italy}
\newcommand{\Ingenieria}{Universidad Nacional de Ingenier{\'\i}a, Lima 25, Per{\'u}}
\newcommand{\INR}{Institute for Nuclear Research of the Russian Academy of Sciences, Moscow 117312, Russia}
\newcommand{\Insubria }{University of Insubria, Via Ravasi, 2, 21100 Varese VA, Italy}
\newcommand{\Iowa}{University of Iowa, Iowa City, IA 52242, USA}
\newcommand{\IowaState}{Iowa State University, Ames, Iowa 50011, USA}
\newcommand{\IPLyon}{Institut de Physique des 2 Infinis de Lyon, 69622 Villeurbanne, France}
\newcommand{\IPM}{Institute for Research in Fundamental Sciences, Tehran, Iran}
\newcommand{\ISTlisboa}{Instituto Superior T{\'e}cnico - IST, Universidade de Lisboa, Portugal}
\newcommand{\Ita}{Instituto Tecnol{\'o}gico de Aeron{\'a}utica, Sao Jose dos Campos, Brazil}
\newcommand{\Iwate}{Iwate University, Morioka, Iwate 020-8551, Japan}
\newcommand{\Jammu}{University of Jammu, Jammu-180006, India}
\newcommand{\Jawaharlal}{Jawaharlal Nehru University, New Delhi 110067, India}
\newcommand{\Jeonbuk}{Jeonbuk National University, Jeonrabuk-do 54896, South Korea}
\newcommand{\JINR}{Joint Institute for Nuclear Research, Dzhelepov Laboratory of Nuclear Problems 6 Joliot-Curie, Dubna, Moscow Region, 141980 RU }
\newcommand{\Jyvaskyla}{University of Jyvaskyla, FI-40014, Finland}
\newcommand{\Kansasstate}{Kansas State University, Manhattan, KS 66506, USA}
\newcommand{\Kavli}{Kavli Institute for the Physics and Mathematics of the Universe, Kashiwa, Chiba 277-8583, Japan}
\newcommand{\KEK}{High Energy Accelerator Research Organization (KEK), Ibaraki, 305-0801, Japan}
\newcommand{\KISTI}{Korea Institute of Science and Technology Information, Daejeon, 34141, South Korea}
\newcommand{\KL}{K L University, Vaddeswaram, Andhra Pradesh 522502, India}
\newcommand{\Kure}{National Institute of Technology, Kure College, Hiroshima, 737-8506, Japan}
\newcommand{\Kyiv}{Taras Shevchenko National University of Kyiv, 01601 Kyiv, Ukraine}
\newcommand{\Lancaster}{Lancaster University, Lancaster LA1 4YB, United Kingdom}
\newcommand{\LawrenceBerkeley}{Lawrence Berkeley National Laboratory, Berkeley, CA 94720, USA}
\newcommand{\LIP}{Laborat{\'o}rio de Instrumenta{\c{c}}{\~a}o e F{\'\i}sica Experimental de Part{\'\i}culas, 1649-003 Lisboa and 3004-516 Coimbra, Portugal}
\newcommand{\Liverpool}{University of Liverpool, L69 7ZE, Liverpool, United Kingdom}
\newcommand{\LosAlmos}{Los Alamos National Laboratory, Los Alamos, NM 87545, USA}
\newcommand{\Louisanastate}{Louisiana State University, Baton Rouge, LA 70803, USA}
\newcommand{\Lucknow}{University of Lucknow, Uttar Pradesh 226007, India}
\newcommand{\Madrid}{Madrid Autonoma University and IFT UAM/CSIC, 28049 Madrid, Spain}
\newcommand{\Mainz}{Johannes Gutenberg-Universit{\"a}t Mainz, 55122 Mainz, Germany}
\newcommand{\Manchester}{University of Manchester, Manchester M13 9PL, United Kingdom}
\newcommand{\Massinsttech}{Massachusetts Institute of Technology, Cambridge, MA 02139, USA}
\newcommand{\Maxplanck}{Max-Planck-Institut, Munich, 80805, Germany}
\newcommand{\Medellin}{University of Medell{\'\i}n, Medell{\'\i}n, 050026 Colombia }
\newcommand{\Michigan}{University of Michigan, Ann Arbor, MI 48109, USA}
\newcommand{\Michiganstate}{Michigan State University, East Lansing, MI 48824, USA}
\newcommand{\MilanoBicocca}{Universit{\`a} del Milano-Bicocca, 20126 Milano, Italy}
\newcommand{\MilanoUniv}{Universit{\`a} degli Studi di Milano, I-20133 Milano, Italy}
\newcommand{\Minnduluth}{University of Minnesota Duluth, Duluth, MN 55812, USA}
\newcommand{\Minntwin}{University of Minnesota Twin Cities, Minneapolis, MN 55455, USA}
\newcommand{\Mississippi}{University of Mississippi, University, MS 38677 USA}
\newcommand{\napoli}{Universit{\`a} degli Studi di Napoli Federico II , 80138 Napoli NA, Italy}
\newcommand{\Newmexico}{University of New Mexico, Albuquerque, NM 87131, USA}
\newcommand{\Niewodniczanski}{H. Niewodnicza{\'n}ski Institute of Nuclear Physics, Polish Academy of Sciences, Cracow, Poland}
\newcommand{\Nikhef}{Nikhef National Institute of Subatomic Physics, 1098 XG Amsterdam, Netherlands}
\newcommand{\Northdakota}{University of North Dakota, Grand Forks, ND 58202-8357, USA}
\newcommand{\Northernillinois}{Northern Illinois University, DeKalb, IL 60115, USA}
\newcommand{\Northwestern}{Northwestern University, Evanston, Il 60208, USA}
\newcommand{\NotreDame}{University of Notre Dame, Notre Dame, IN 46556, USA}
\newcommand{\NoviSad}{University of Novi Sad, 21102 Novi Sad, Serbia}
\newcommand{\Occidental}{Occidental College, Los Angeles, CA  90041}
\newcommand{\Ohiostate}{Ohio State University, Columbus, OH 43210, USA}
\newcommand{\OregonState}{Oregon State University, Corvallis, OR 97331, USA}
\newcommand{\Oxford}{University of Oxford, Oxford, OX1 3RH, United Kingdom}
\newcommand{\PacificNorthwest}{Pacific Northwest National Laboratory, Richland, WA 99352, USA}
\newcommand{\Padova}{Universt{\`a} degli Studi di Padova, I-35131 Padova, Italy}
\newcommand{\Panjab}{Panjab University, Chandigarh, 160014 U.T., India}
\newcommand{\Parissaclay}{Universit{\'e} Paris-Saclay, CNRS/IN2P3, IJCLab, 91405 Orsay, France}
\newcommand{\Parisuniversite}{Universit{\'e} Paris Cit{\'e}, CNRS, Astroparticule et Cosmologie, Paris, France}
\newcommand{\Parma}{University of Parma,  43121 Parma PR, Italy}
\newcommand{\Pavia}{Universit{\`a} degli Studi di Pavia, 27100 Pavia PV, Italy}
\newcommand{\Penn}{University of Pennsylvania, Philadelphia, PA 19104, USA}
\newcommand{\PennState}{Pennsylvania State University, University Park, PA 16802, USA}
\newcommand{\PhysicalResearchLaboratory}{Physical Research Laboratory, Ahmedabad 380 009, India}
\newcommand{\Pisa}{Universit{\`a} di Pisa, I-56127 Pisa, Italy}
\newcommand{\Pitt}{University of Pittsburgh, Pittsburgh, PA 15260, USA}
\newcommand{\Pontificia}{Pontificia Universidad Cat{\'o}lica del Per{\'u}, Lima, Per{\'u}}
\newcommand{\PuertoRico}{University of Puerto Rico, Mayaguez 00681, Puerto Rico, USA}
\newcommand{\Punjab}{Punjab Agricultural University, Ludhiana 141004, India}
\newcommand{\QMUL}{Queen Mary University of London, London E1 4NS, United Kingdom }
\newcommand{\Radboud}{Radboud University, NL-6525 AJ Nijmegen, Netherlands}
\newcommand{\Rochester}{University of Rochester, Rochester, NY 14627, USA}
\newcommand{\Royalholloway}{Royal Holloway College London, TW20 0EX, United Kingdom}
\newcommand{\Rutgers}{Rutgers University, Piscataway, NJ, 08854, USA}
\newcommand{\Rutherford}{STFC Rutherford Appleton Laboratory, Didcot OX11 0QX, United Kingdom}
\newcommand{\Salento}{Universit{\`a} del Salento, 73100 Lecce, Italy}
\newcommand{\Sanjosestate}{San Jose State University, San Jos{\'e}, CA 95192-0106, USA}
\newcommand{\Sapienza}{Sapienza University of Rome, 00185 Roma RM, Italy}
\newcommand{\SergioArboleda}{Universidad Sergio Arboleda, 11022 Bogot{\'a}, Colombia}
\newcommand{\Sheffield}{University of Sheffield, Sheffield S3 7RH, United Kingdom}
\newcommand{\SLAC}{SLAC National Accelerator Laboratory, Menlo Park, CA 94025, USA}
\newcommand{\Southcarolina}{University of South Carolina, Columbia, SC 29208, USA}
\newcommand{\SouthDakotaSchool}{South Dakota School of Mines and Technology, Rapid City, SD 57701, USA}
\newcommand{\SouthDakotaState}{South Dakota State University, Brookings, SD 57007, USA}
\newcommand{\SouthernMethodist}{Southern Methodist University, Dallas, TX 75275, USA}
\newcommand{\StonyBrook}{Stony Brook University, SUNY, Stony Brook, NY 11794, USA}
\newcommand{\Sunyatsen}{Sun Yat-Sen University, Guangzhou, 510275, China}
\newcommand{\SURF}{Sanford Underground Research Facility, Lead, SD, 57754, USA}
\newcommand{\Sussex}{University of Sussex, Brighton, BN1 9RH, United Kingdom}
\newcommand{\Syracuse}{Syracuse University, Syracuse, NY 13244, USA}
\newcommand{\Tecnologica }{Universidade Tecnol{\'o}gica Federal do Paran{\'a}, Curitiba, Brazil}
\newcommand{\TexasAMcollege}{Texas A{\&}M University, College Station, Texas 77840}
\newcommand{\TexasAMcorpuscristi}{Texas A{\&}M University - Corpus Christi, Corpus Christi, TX 78412, USA}
\newcommand{\TexasArlington}{University of Texas at Arlington, Arlington, TX 76019, USA}
\newcommand{\Texasaustin}{University of Texas at Austin, Austin, TX 78712, USA}
\newcommand{\Toronto}{University of Toronto, Toronto, Ontario M5S 1A1, Canada}
\newcommand{\Tufts}{Tufts University, Medford, MA 02155, USA}
\newcommand{\Unifesp}{Universidade Federal de S{\~a}o Paulo, 09913-030, S{\~a}o Paulo, Brazil}
\newcommand{\UNIST}{Ulsan National Institute of Science and Technology, Ulsan 689-798, South Korea}
\newcommand{\UniversityCollegeLondon}{University College London, London, WC1E 6BT, United Kingdom}
\newcommand{\ValleyCity}{Valley City State University, Valley City, ND 58072, USA}
\newcommand{\VariableEnergy}{Variable Energy Cyclotron Centre, 700 064 West Bengal, India}
\newcommand{\VirginiaTech}{Virginia Tech, Blacksburg, VA 24060, USA}
\newcommand{\Warsaw}{University of Warsaw, 02-093 Warsaw, Poland}
\newcommand{\Warwick}{University of Warwick, Coventry CV4 7AL, United Kingdom}
\newcommand{\Wellesley}{Wellesley College, Wellesley, MA 02481, USA}
\newcommand{\Wichita}{Wichita State University, Wichita, KS 67260, USA}
\newcommand{\WilliamMary}{William and Mary, Williamsburg, VA 23187, USA}
\newcommand{\Wisconsin}{University of Wisconsin Madison, Madison, WI 53706, USA}
\newcommand{\Yale}{Yale University, New Haven, CT 06520, USA}
\newcommand{\Yerevan}{Yerevan Institute for Theoretical Physics and Modeling, Yerevan 0036, Armenia}
\newcommand{\York}{York University, Toronto M3J 1P3, Canada}
\affiliation{\Abilene}
\affiliation{\Albanysuny}
\affiliation{\Amsterdam}
\affiliation{\Antalya}
\affiliation{\Antananarivo}
\affiliation{\AntonioNarino}
\affiliation{\Argonne}
\affiliation{\Arizona}
\affiliation{\Asuncion}
\affiliation{\Athens}
\affiliation{\Atlantico}
\affiliation{\Augustana}
\affiliation{\Basel}
\affiliation{\Bern}
\affiliation{\Beykent}
\affiliation{\Birmingham}
\affiliation{\BolognaUniversity}
\affiliation{\Boston}
\affiliation{\Bristol}
\affiliation{\Brookhaven}
\affiliation{\Bucharest}
\affiliation{\CalBerkeley}
\affiliation{\CalDavis}
\affiliation{\CalIrvine}
\affiliation{\CalLosangeles}
\affiliation{\CalRiverside}
\affiliation{\CalSantabarbara}
\affiliation{\Caltech}
\affiliation{\Cambridge}
\affiliation{\Campinas}
\affiliation{\CataniaUniversitadi}
\affiliation{\Catolica}
\affiliation{\CBPF}
\affiliation{\CEASaclay}
\affiliation{\CERN}
\affiliation{\Charles}
\affiliation{\Chicago}
\affiliation{\ChungAng}
\affiliation{\CIEMAT}
\affiliation{\Cincinnati}
\affiliation{\Cinvestav}
\affiliation{\Colima}
\affiliation{\ColoradoBoulder}
\affiliation{\ColoradoState}
\affiliation{\Columbia}
\affiliation{\conida}
\affiliation{\Cti}
\affiliation{\CUSB}
\affiliation{\CzechAcademyofSciences}
\affiliation{\CzechTechnical}
\affiliation{\DakotaState}
\affiliation{\Dallas}
\affiliation{\DannecyleVieux}
\affiliation{\Daresbury}
\affiliation{\Drexel}
\affiliation{\Duke}
\affiliation{\Durham}
\affiliation{\Edinburgh}
\affiliation{\EIA}
\affiliation{\ETH}
\affiliation{\Eotvos}
\affiliation{\FCULport}
\affiliation{\FederaldeAlfenas}
\affiliation{\FederaldeGoias}
\affiliation{\FederaldoABC}
\affiliation{\FederaldoRio}
\affiliation{\Fermi}
\affiliation{\Ferrarauniv}
\affiliation{\Florida}
\affiliation{\Floridastate}
\affiliation{\Fluminense}
\affiliation{\Genova}
\affiliation{\Georgian}
\affiliation{\Granada}
\affiliation{\GranSasso}
\affiliation{\GranSassoLab}
\affiliation{\Grenoble}
\affiliation{\Guanajuato}
\affiliation{\Harish}
\affiliation{\Harvard}
\affiliation{\Hawaii}
\affiliation{\Houston}
\affiliation{\Hyderabad}
\affiliation{\Idaho}
\affiliation{\IFAE}
\affiliation{\IFIC}
\affiliation{\IGFAE}
\affiliation{\Illinoisinstitute}
\affiliation{\Imperial}
\affiliation{\IndGuwahati}
\affiliation{\IndHyderabad}
\affiliation{\Indiana}
\affiliation{\INFNBologna}
\affiliation{\INFNCatania}
\affiliation{\INFNFerrara}
\affiliation{\INFNGenova}
\affiliation{\INFNLecce}
\affiliation{\INFNMilanBicocca}
\affiliation{\INFNMilano}
\affiliation{\INFNNapoli}
\affiliation{\INFNPadova}
\affiliation{\INFNPavia}
\affiliation{\INFNPisa}
\affiliation{\INFNRoma}
\affiliation{\INFNSud}
\affiliation{\Ingenieria}
\affiliation{\INR}
\affiliation{\Insubria }
\affiliation{\Iowa}
\affiliation{\IowaState}
\affiliation{\IPLyon}
\affiliation{\IPM}
\affiliation{\ISTlisboa}
\affiliation{\Ita}
\affiliation{\Iwate}
\affiliation{\Jammu}
\affiliation{\Jawaharlal}
\affiliation{\Jeonbuk}
\affiliation{\JINR}
\affiliation{\Jyvaskyla}
\affiliation{\Kansasstate}
\affiliation{\Kavli}
\affiliation{\KEK}
\affiliation{\KISTI}
\affiliation{\KL}
\affiliation{\Kure}
\affiliation{\Kyiv}
\affiliation{\Lancaster}
\affiliation{\LawrenceBerkeley}
\affiliation{\LIP}
\affiliation{\Liverpool}
\affiliation{\LosAlmos}
\affiliation{\Louisanastate}
\affiliation{\Lucknow}
\affiliation{\Madrid}
\affiliation{\Mainz}
\affiliation{\Manchester}
\affiliation{\Massinsttech}
\affiliation{\Maxplanck}
\affiliation{\Medellin}
\affiliation{\Michigan}
\affiliation{\Michiganstate}
\affiliation{\MilanoBicocca}
\affiliation{\MilanoUniv}
\affiliation{\Minnduluth}
\affiliation{\Minntwin}
\affiliation{\Mississippi}
\affiliation{\napoli}
\affiliation{\Newmexico}
\affiliation{\Niewodniczanski}
\affiliation{\Nikhef}
\affiliation{\Northdakota}
\affiliation{\Northernillinois}
\affiliation{\Northwestern}
\affiliation{\NotreDame}
\affiliation{\NoviSad}
\affiliation{\Occidental}
\affiliation{\Ohiostate}
\affiliation{\OregonState}
\affiliation{\Oxford}
\affiliation{\PacificNorthwest}
\affiliation{\Padova}
\affiliation{\Panjab}
\affiliation{\Parissaclay}
\affiliation{\Parisuniversite}
\affiliation{\Parma}
\affiliation{\Pavia}
\affiliation{\Penn}
\affiliation{\PennState}
\affiliation{\PhysicalResearchLaboratory}
\affiliation{\Pisa}
\affiliation{\Pitt}
\affiliation{\Pontificia}
\affiliation{\PuertoRico}
\affiliation{\Punjab}
\affiliation{\QMUL}
\affiliation{\Radboud}
\affiliation{\Rochester}
\affiliation{\Royalholloway}
\affiliation{\Rutgers}
\affiliation{\Rutherford}
\affiliation{\Salento}
\affiliation{\Sanjosestate}
\affiliation{\Sapienza}
\affiliation{\SergioArboleda}
\affiliation{\Sheffield}
\affiliation{\SLAC}
\affiliation{\Southcarolina}
\affiliation{\SouthDakotaSchool}
\affiliation{\SouthDakotaState}
\affiliation{\SouthernMethodist}
\affiliation{\StonyBrook}
\affiliation{\Sunyatsen}
\affiliation{\SURF}
\affiliation{\Sussex}
\affiliation{\Syracuse}
\affiliation{\Tecnologica }
\affiliation{\TexasAMcollege}
\affiliation{\TexasAMcorpuscristi}
\affiliation{\TexasArlington}
\affiliation{\Texasaustin}
\affiliation{\Toronto}
\affiliation{\Tufts}
\affiliation{\Unifesp}
\affiliation{\UNIST}
\affiliation{\UniversityCollegeLondon}
\affiliation{\ValleyCity}
\affiliation{\VariableEnergy}
\affiliation{\VirginiaTech}
\affiliation{\Warsaw}
\affiliation{\Warwick}
\affiliation{\Wellesley}
\affiliation{\Wichita}
\affiliation{\WilliamMary}
\affiliation{\Wisconsin}
\affiliation{\Yale}
\affiliation{\Yerevan}
\affiliation{\York}
\author{A.~Abed Abud} \affiliation{\CERN}
\author{B.~Abi} \affiliation{\Oxford}
\author{R.~Acciarri} \affiliation{\Fermi}
\author{M.~A.~Acero} \affiliation{\Atlantico}
\author{M.~R.~Adames} \affiliation{\Tecnologica }
\author{G.~Adamov} \affiliation{\Georgian}
\author{M.~Adamowski} \affiliation{\Fermi}
\author{D.~Adams} \affiliation{\Brookhaven}
\author{M.~Adinolfi} \affiliation{\Bristol}
\author{C.~Adriano} \affiliation{\Campinas}
% ...
\collaboration{The DUNE Collaboration}
\noaffiliation



\date{\today}

\begin{abstract}
Measurements of electrons from $\nu_e$ interactions are crucial for the Deep Underground Neutrino Experiment (DUNE) neutrino oscillation program, as well as searches for physics beyond the standard model, supernova neutrino detection, and solar neutrino measurements. This article describes the selection and reconstruction of low-energy (Michel) electrons in the ProtoDUNE-SP detector. ProtoDUNE-SP is one of the prototypes for the DUNE far detector, built and operated at CERN as a charged particle test beam experiment.
A sample of low-energy electrons produced by the decay of cosmic muons is selected with a purity of 95\%.
This sample is used to calibrate the low-energy electron energy scale with two techniques.
An electron energy calibration based on a cosmic ray muon sample uses calibration constants derived from measured and simulated cosmic ray muon events. Another calibration technique makes use of the theoretically well-understood Michel electron energy spectrum to convert reconstructed charge to electron energy.
In addition, the effects of detector response to low-energy electron energy scale and its resolution including readout electronics threshold effects are quantified.
Finally, the relation between the theoretical and reconstructed low-energy electron energy spectrum is derived and the energy resolution is characterized. The low-energy electron selection presented here accounts for about 75\% of the total electron deposited energy. After the addition of lost energy using a Monte Carlo simulation, the energy resolution improves from about 40\% to 25\% at 50~MeV. These results are used to validate the expected capabilities of the DUNE far detector to reconstruct low-energy electrons.
\end{abstract}


\maketitle  
\section{Introduction}

Discoveries over the past half-century have positioned neutrinos, one of the most abundant matter particles in the Universe, at the center stage of fundamental physics. Neutrinos are now being studied to answer open questions about the nature of matter and the evolution of the Universe.  In particular, the measurement of $CP$ violation in the lepton sector~\cite{cp-violation-davidson, cp-violation-fukugita} will help probe the possibility that early-Universe $CP$ violation involving leptons might have led to the present dominance of matter over antimatter. DUNE~\cite{dune_osc_paper, dune-physics-TDR} is a next-generation long-baseline accelerator neutrino experiment, designed to be sensitive to neutrino oscillations.
The DUNE experiment will consist of a far detector~\cite{dune-intro-TDR} to be located about 1.5~km underground at the Sanford Underground Research Facility (SURF) in South Dakota, USA, at a distance of 1300~km from Fermilab, and a near detector~\cite{dune-nd-cdr} to be located at Fermilab. 
DUNE uses liquid argon time projection chamber (LArTPC) technology, which permits the reconstruction of neutrino interactions with mm-scale precision.  
$CP$ violation will be tested in $\nu_{\mu} \rightarrow \nu_e$ oscillations and the corresponding anti-neutrino channel, which are sensitive
to the CP-violating phase and the neutrino mass ordering~\cite{mass_hierarchy}. 
In addition, the large underground LArTPC detectors planned for DUNE will enable a rich physics program beyond the accelerator-based neutrino oscillation program, including searches beyond the standard model~\cite{dune_bsm}, supernova neutrino detection~\cite{dune_sn}, and solar neutrino measurements~\cite{dune_solar}.

To achieve the planned DUNE physics program, it is critically important to accurately reconstruct the energies of electrons and positrons originating from MeV-scale solar and supernova burst $\nu_e$'s as well as GeV-scale neutrinos from the Long-Baseline Neutrino Facility beam.
Calorimetric energy reconstruction requires efficient charge collection, calibration corrections to account for liquid argon impurities and electronics response, and a recombination correction to account for charge loss due to  electron-ion recombination.
The goal of this article is to demonstrate the capability to reconstruct low-energy electrons in the single-phase ProtoDUNE (ProtoDUNE-SP)~\cite{protodune-paper} LArTPC. This work presents techniques and results on the selection and energy reconstruction of the low-energy (Michel) electrons~\cite{original_Michel}, originating from the decay at rest of cosmic ray muons. With a well-understood energy spectrum, these low-energy electrons are ideal for evaluating the electron selection and energy reconstruction in ProtoDUNE-SP and demonstrating the capability of the DUNE far detector to identify and reconstruct these low-energy electron events.
Although there are other studies of low-energy electrons in LArTPCs~\cite{MicroBooNE_Michel, Michel_lariat, Michel_ArgoNeuT, MicroBooNE_stream, Snowmass_paper}, the unique features of this study include the data-driven determination of the recombination correction,
evaluation of the lost energy due to the TPC readout threshold,
a comparison of the electron energy calibration based on muon-derived calibration corrections with that based on the Michel electron true energy spectrum, and a characterization of the electron energy resolution.

\section{DUNE first far detector and its prototype 
}

Central to the realization of the DUNE physics program is the construction and operation of LArTPC detectors that combine a many-kiloton fiducial mass necessary for rare-event searches with the ability to image those events with mm-scale spatial resolution, providing the capability to identify the signatures of the physics processes of interest.
The DUNE far detector will consist of four detector modules, each with an equivalent LAr fiducial mass of 10~kt, installed approximately 1.5~km underground.
Each LArTPC will be installed inside a cryostat of internal dimensions 15.1~m~(w) $\times$ 14~m~(h) $\times$ 62~m~(l) containing a total LAr mass of about 17.5~kt. Charged particles passing through the TPC ionize the argon, and the ionization electrons 
drift to the anode planes under the influence of an applied electric field.

DUNE is actively developing two LArTPC technologies: a horizontal-drift (HD) LArTPC in which the ionization electrons drift horizontally between a vertical cathode and anode planes, and a vertical-drift LArTPC, in which the ionization electrons drift vertically between a horizontal cathode and
anode planes. The focus of this article is on the HD LArTPC~\cite{dune-fd-sp-TDR} technology as the first DUNE far detector module will be based on this technology.
\begin{figure}[!htpb]
\begin{minipage}{.5\textwidth}
\centering
\includegraphics[width=0.8\linewidth]{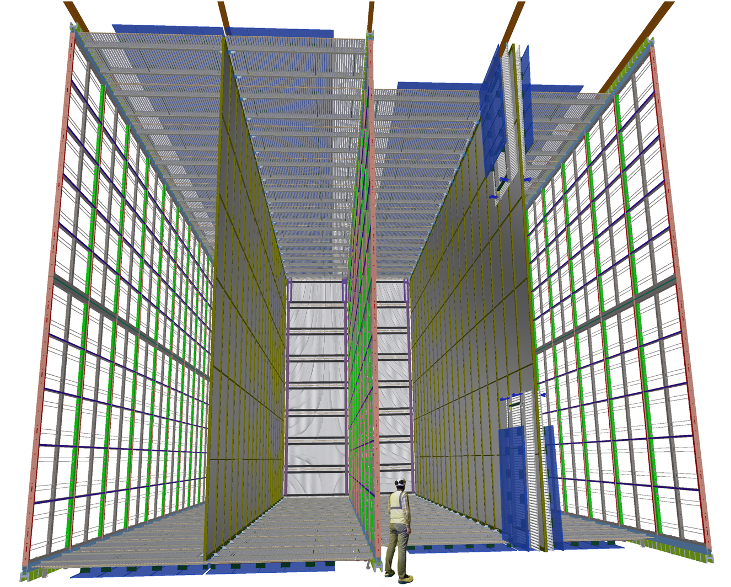}
\end{minipage}
\begin{minipage}{.5\textwidth}
\includegraphics[width=0.8\linewidth]{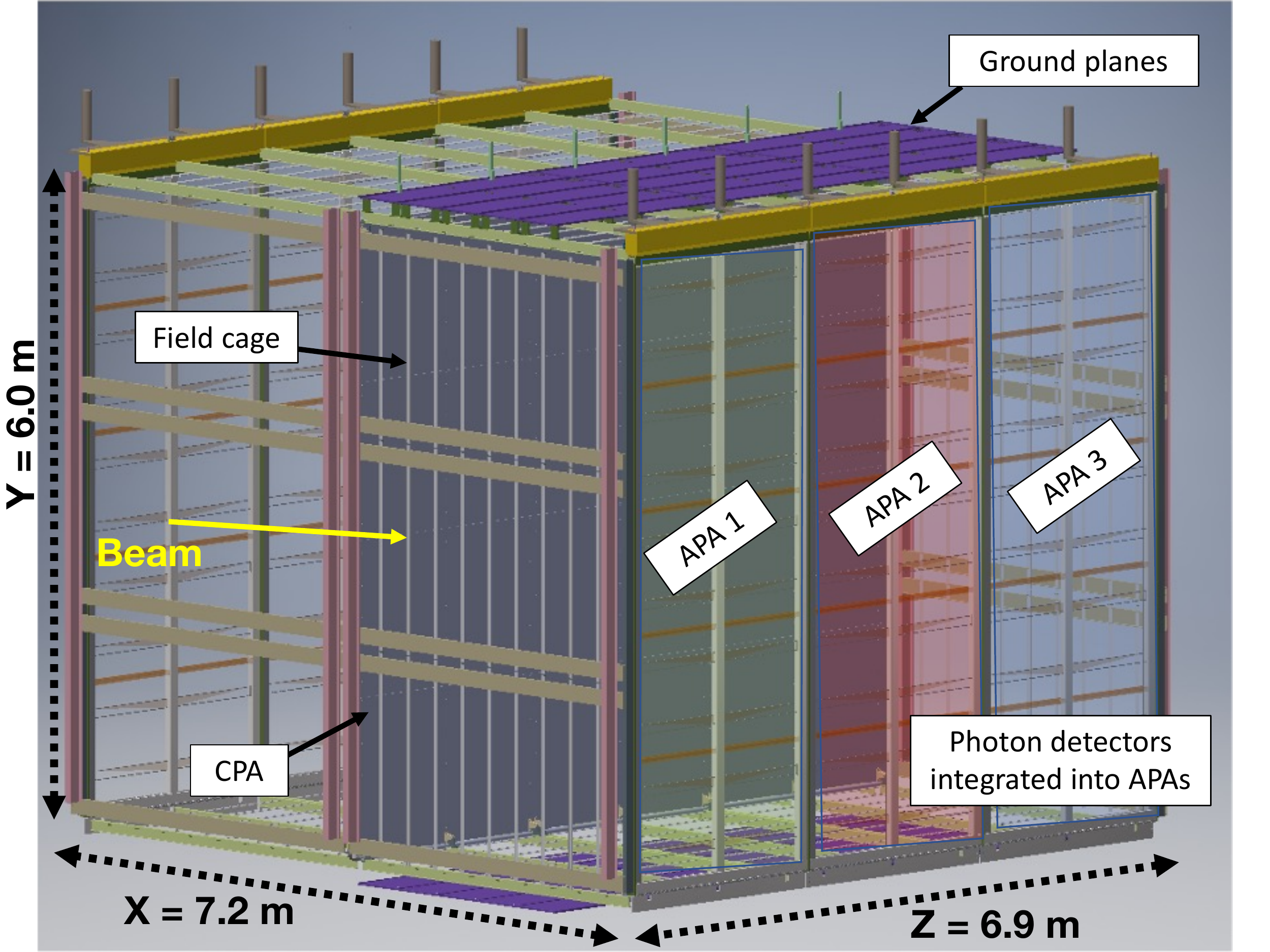}
\end{minipage}\par\medskip
\caption{Configuration of the 10~kt DUNE far detector horizontal drift module (top); Configuration of ProtoDUNE-SP LArTPC (bottom).}
\label{fig:dune_fd_sp}
\end{figure}

Figure~\ref{fig:dune_fd_sp} (top) shows the configuration of a DUNE HD module. Each of the four LAr drift volumes is subjected to an electric field of 500~V/cm~\cite{electric_field_value}, corresponding to a Cathode Plane Assembly (CPA) high voltage of -180~kV relative to the anode, which will be grounded.
The pattern of ionization collected on the grid of anode wires enables reconstruction in the two coordinates perpendicular to the drift direction. Novel 
photon detectors (PDs) called X-ARAPUCAs~\cite{arapuca} will be placed 
behind the Anode Plane Assembly (APA) collection wire planes. 
The PDs are used to provide a timestamp of the interaction, thus giving an estimate of the drift distances traveled by the ionization electrons to reconstruct the third event coordinate.

The DUNE collaboration has constructed and operated a large horizontal drift prototype detector, known as 
ProtoDUNE-SP. The detector has been assembled and tested at the CERN Neutrino Platform~\cite{cern-nu-platform}.
ProtoDUNE-SP was operated from 2018 to 2020 and its large samples of high-quality beam data have been used to demonstrate the effectiveness of the single-phase far detector design. Results on the performance of the ProtoDUNE-SP liquid argon TPC in the test beam can be found in reference~\cite{perf_paper} including noise and gain measurements, $dE/dx$ calibration for muons, protons, pions and electrons, drift electron lifetime measurements, and photon detector noise, signal sensitivity and time resolution measurements. The measured values meet or exceed the specifications for the DUNE far detector.
Figure~\ref{fig:dune_fd_sp} (bottom) shows the components of the ProtoDUNE-SP LArTPC, which is approximately one-twentieth the size of the planned far detector HD module but uses anode and cathode components identical in size to those of the full-scale module. ProtoDUNE-SP has the same 3.6~m maximum drift length as the full far detector HD module. It consists of two drift volumes with a common central cathode surrounded 
by two anode planes and a field cage that surrounds the entire active volume. The active volume is 6~m high (y-coordinate), 7.2~m wide (x-coordinate, along the drift direction) and 7~m deep (z-coordinate, along the beam direction). 

Each anode plane consists of three adjacent APAs that are each 6~m high by 2.3~m wide. The wire planes and their wire orientations are the U layer ($+35.7^\circ$ from vertical, also called the first induction plane), the V layer ($-35.7^\circ$ from vertical, also called the second induction plane), and the X layer (vertical, also called the collection plane). Each successive wire plane is built 4.75 mm above the previous layer.
As they drift, ionization electrons first pass the induction planes and then are collected on the collection plane.
The U and V plane wires are wrapped around the APA frame (and hence see the charge arriving from both sides of the APA) while each side has a separate X layer, as sketched in Figure 3 of~\cite{perf_paper}. The distance between two consecutive wires in the same layer, also known as wire pitch, is $4.67$~mm for U and V layers, and $4.79$~mm for X layer wires. Signals from the wires of each APA are read out via a total of 2560 electronic channels.

Uniformity of the electric field is provided by the surrounding field cage. The cold electronics mounted onto the APA frame, and thus immersed in LAr, amplify and continuously digitize the induced signals on the sense wires at 2 MHz during the entire data-taking period, and transmit these waveforms to the Data Acquisition system. 
The modular PD system is integrated into the APAs, as further described in~\cite{perf_paper}. The PD was not used in the analysis described here.

\section{Electrons in LArTPCs}

For the DUNE physics program it is critical to understand the far detector response to electromagnetic showers since DUNE will measure electrons produced in $\nu_e$ interactions, where the $\nu_e$ are from $\nu_\mu$ oscillations, the Sun and possibly supernova explosions. In addition, DUNE will search for proton decay signatures, as event identification may proceed via the detection of a low-energy electron.
ProtoDUNE-SP has collected data samples of test-beam electrons and data samples of electrons from cosmic ray muon decays~\cite{perf_paper}.
Data from ProtoDUNE-SP beam runs with 1 GeV/c beam momentum, including a sample of beam positrons, were used for the initial classification of track- and shower-like energy deposits using a convolutional neural network technique~\cite{protodune_CNN}.
Studies of electron selection and identification in ProtoDUNE-SP TPC lead to a more accurate understanding of the calorimetric response to electrons
and offer an opportunity for a precise understanding of the electron energy resolution parameters for electron neutrino reconstruction in future DUNE far detectors.
This work focuses on studies of the ProtoDUNE-SP LArTPC response to low-energy electrons. 

As the electrons propagate in the LAr, they deposit energy either through ionization or through radiative losses (bremsstrahlung). The energy loss via ionization is continuous and results in track-like topologies.
Radiative losses are also present at all electron energies 
leading to the production of electromagnetic shower cascades of secondary electrons and photons. Bremsstrahlung photons may Compton scatter or convert to $e^+e^-$ pairs, resulting in signatures with secondary energy deposits disconnected from the primary ionization tracks. The typical attenuation length for photons in liquid argon in the energy range of interest for Michel electrons is 20$-$30~cm~\cite{attenuation_length}.
The event reconstruction takes into account the charge released by both primary particle ionization and radiative processes.

\section{Selection of Stopping Muons and Michel Electrons}

The generation of cosmic ray muons is performed with CORSIKA~v7.4~\cite{Corsika}, while the simulation of particle propagation and interaction in ProtoDUNE-SP is performed by Geant4~v4.10.3 by using the QGSP BERT physics list~\cite{GEANT} with the detector response described within LArSoft~\cite{LArsoft}.
In all ProtoDUNE simulations, the delta-ray threshold (and the electron transport threshold) is set to 455~keV (corresponding to an electron range of about 1.5~mm)~\cite{GEANT_webpage}.
All $\mu^+$ decay into Michel positrons, whereas only $25\%$ of $\mu^-$ undergo decay to Michel electrons since the other $75\%$ are captured by the argon atoms inside the TPC. Therefore, the Michel electron sample described in this analysis includes both electrons and positrons. In this article, ``electrons" refers to both electrons and positrons unless indicated otherwise.

The reconstruction of charged particles in the ProtoDUNE-SP LArTPC follows the technique described elsewhere~\cite{perf_paper}, and in this section the procedure is briefly described.
The TPC readout electronics collect a waveform that represents the current on the APA wire as a function of time. Each waveform is processed in an offline data processing chain to produce a collection of 
ionization charge deposition
arrival times and charge integrals at each readout wire. 
Signal processing starts with a deconvolution of measured charge from signals induced by the drifting ionization electrons, followed by noise removal. 
In order to make use of deconvolved waveforms to reconstruct individual events, it is necessary to apply three-dimensional (3D) hit finding and pattern recognition algorithms.
The 3D-hit (called “hit” from now on) is an ionization charge released in space and time by through-going charged particles and detected by three layers of anode plane wires, and collected by a collection plane wire alone in the analysis described here. A collection of hits is merged together to form a particle track or a shower that belongs to an event.
The hit finding algorithm searches for candidate hits based on charge deposits in the waveform on a single wire as a function of time, and fits them to a Gaussian shape. Pattern recognition and event reconstruction are performed by the Pandora software package~\cite{pand_package}, which is a collection of reconstruction algorithms that focus on specific hit topology patterns. The first step in the reconstruction procedure is the two-dimensional clustering of observed charge pulses in each of the three detector readout planes separately. 
In the second step, sets of two-dimensional clusters are matched between the three views
to produce 3D hits and to create particle interaction hierarchies.
As described in~\cite{perf_paper}, one important feature of the cosmic ray reconstruction step is the “stitching” of tracks across the boundaries between neighboring drift volumes bounded by a CPA or an APA. 
In the analysis performed here the stitching procedure is applied when two 3D clusters are reconstructed in neighboring drift volumes with consistent direction vectors and an equal but opposite shift in the drift direction from the CPA. These two clusters present segments of a single muon track that is penetrated through the CPA. When the clusters are shifted toward each other as expressed in time-tick units (1~time~tick = 500~ns), a single muon track of two initially separate tracks is produced with a known absolute position and time ($T_0$) relative to the trigger time~\cite{perf_paper}. 

\begin{figure}[!htpb]
\begin{adjustwidth}{-2cm}{-2cm}
\centering
\begin{minipage}{.5\textwidth}
\subfloat{\includegraphics[height=5.0cm, width=1.0\textwidth]{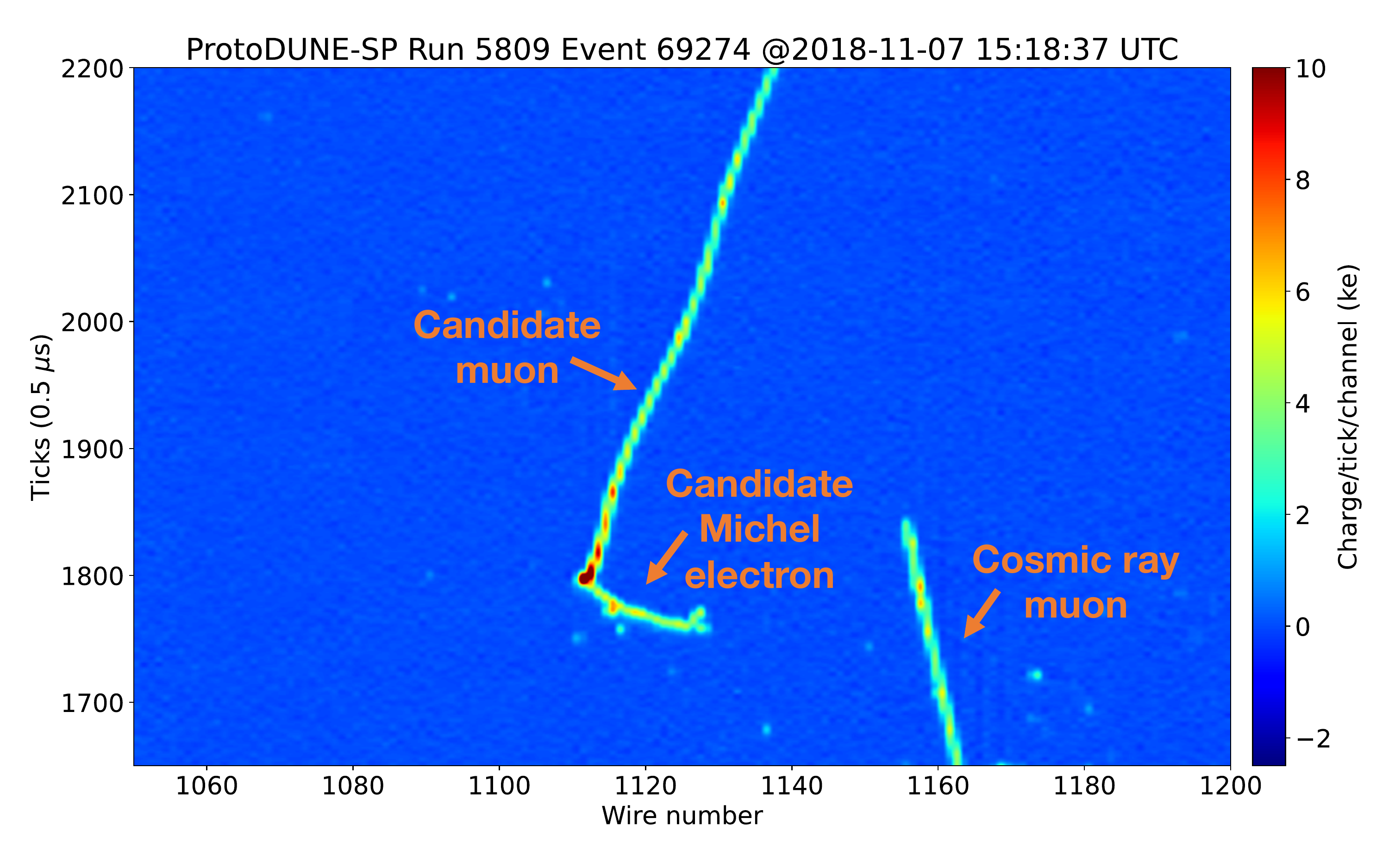}}
\end{minipage}
\begin{minipage}{.5\textwidth}
\subfloat{\includegraphics[height=5.0cm, width=1.0\textwidth]{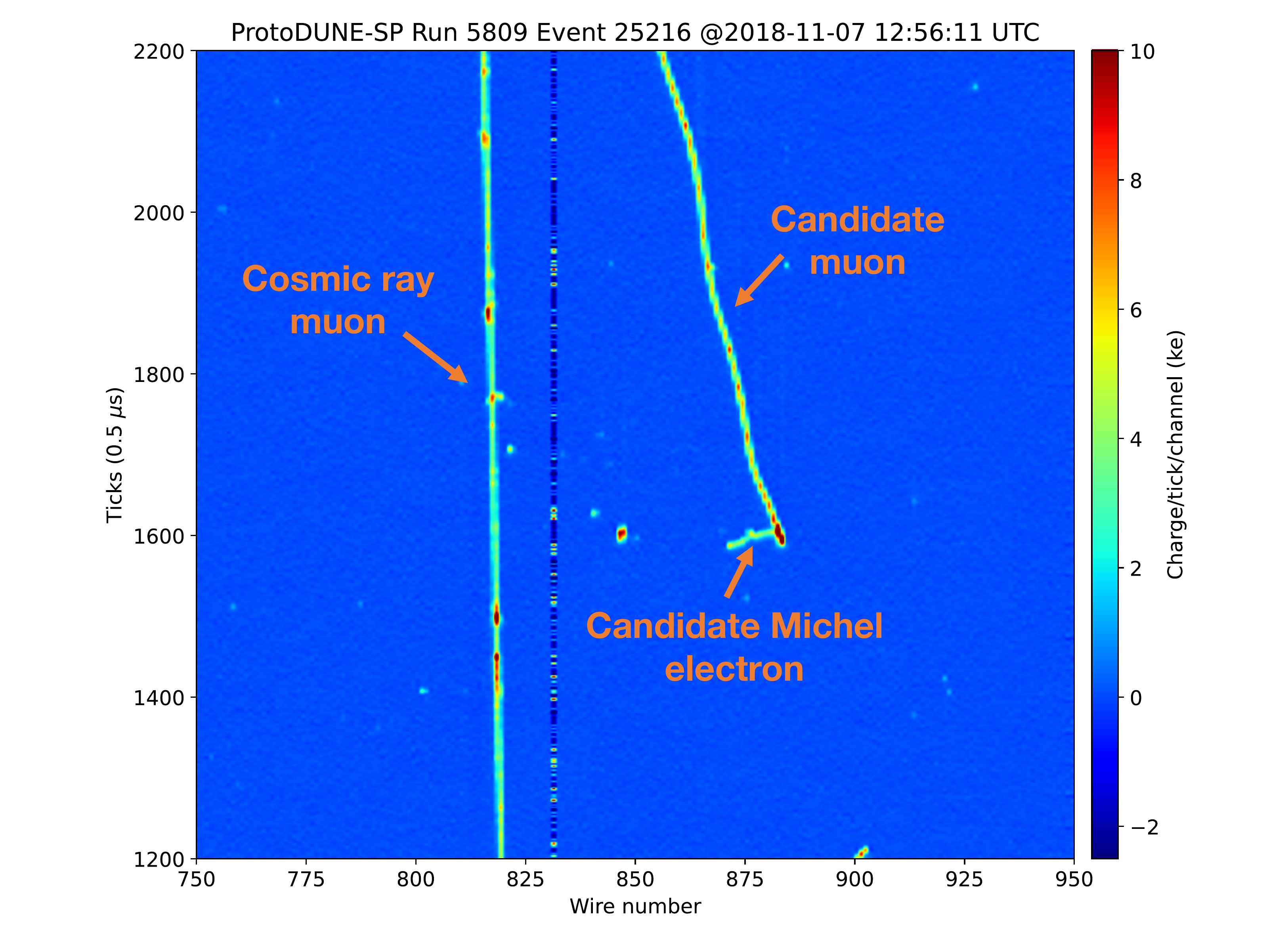}}
\end{minipage}
\caption{Two Michel electron candidates observed in the ProtoDUNE-SP data. The parent muons enter the images from the top before stopping and decaying.}
\label{fig:michel_e_event_display}
\end{adjustwidth}
\end{figure}

The reconstruction of electrons below 50~MeV is very different from the reconstruction of GeV-scale electromagnetic showers~\cite{reconstruction_instruments}. For this reason, a dedicated algorithm has been developed to reconstruct and identify the Michel electrons presented in this study.
Figure~\ref{fig:michel_e_event_display} shows two Michel electron candidate events from ProtoDUNE-SP data, with muons entering from
the top.

The event selection starts by searching for a candidate muon that decayed to an electron. A set of conditions is initially applied to ensure a high quality muon track candidate. 
Finally, additional selection criteria are implemented to make sure that a Michel electron candidate is identified around the end position of the candidate muon by selecting and summing up charge hits that represent the Michel electron. While all three anode planes are used for track reconstruction, the collection plane provides the best signal-to-noise performance and charge resolution~\cite{perf_paper}. Therefore, only the collection plane charge is used to reconstruct the electron energy.  

\subsection{Muon Track Selection}\label{muon_track_selection}

\begin{enumerate}[(i)]
\item Only the $T_{0}$-tagged candidate muon tracks are selected from muon tracks reconstructed by Pandora. These are the tracks that cross the cathode or anode plane boundaries and the two pieces of the track from the two volumes help determine the correct end position of the track in the drift direction. The fraction of tracks having a $T_{0}$ assigned to them is 2\% from the data sample. Since this requirement selects most of the events and the corresponding charge coming from locations farther away from the anode in ProtoDUNE-SP, it is expected that the DUNE far detector will perform equally or better in terms of charge reconstruction. This is because there will be less ionization charge attenuation as events, on average, will be closer to the anode plane wires and that may slightly affect the energy reconstruction. 

\item Selected tracks are required to have one reconstructed endpoint within 30~cm from one or more of the detector boundaries. The cut is applied to all six faces of the detector. This step improves the selection of cosmic ray muon candidates entering the detector. By requiring this, the next steps in the selection can focus on the other end of the track to search for the Michel electron signatures. 

\item Only the muons that stop within the detector fiducial volume are considered. The fiducial volume is a rectangular volume shaped as follows: the boundary from the anode planes is $51\text{~cm}$, the boundary from the upstream and downstream ends is $80\text{~cm}$, and the boundary from the top and bottom of the TPC is $43\text{~cm}$ and $80\text{~cm}$, respectively. These values are obtained from an optimization based on Monte Carlo (MC) simulation.
This step specifies the end of the contained track from which evidence of Michel electrons can be sought.

\item Muon tracks that stop within a region that is close to a boundary between two adjacent APAs ($\sim$10~cm from each APA side) are removed. This cut removes all those tracks that appear to stop in the gaps between two APA planes. 

\item Broken tracks, for which the reconstruction algorithm does not connect track segments correctly at detector boundaries or anode gaps, are removed from further analysis.
In order to reject broken candidate muon tracks, the algorithm looks for any additional track that starts within $<30$~cm of the reconstructed end position of the candidate muon track, and is nearly parallel ($<14^\circ$ or $>165^\circ$)  with respect to the candidate muon track. If this condition is satisfied, the candidate muon is removed from the event selection. 

\item It is required that candidate muon tracks are at least $75$~cm long~\cite{NOvA_seasonal} i.e. those that have crossed the cathode with track segments reconstructed in both drift volumes. Since cosmic muons generally have long track lengths, this cut improves the quality of the candidate cosmic muon track reconstruction.

\item  Every reconstructed hit is associated with a time counted in ticks, known as the hit time with respect to $T_{0}$. The peak of the reconstructed hit time distribution is known as the hit peak time. For every track, a cut is placed on the value of the minimum and maximum hit peak time. Only those candidate muon tracks that have a minimum hit peak time to be $>200$ time ticks and a maximum hit peak time to be $<5800$ time ticks are kept. The peak time cuts ensure that the candidate muon is contained within the event readout window. 

\end{enumerate}

About 28\% of the $T_{0}$-tagged muons satisfy the above selection criteria, and simulation studies indicate that the selected muon sample has a purity of 99.7\%. The purity here corresponds to the fraction of the true muons out of all the selected tracks. The determination of the selection cut values for different quantities is based on the MC simulation studies for which the maximum sample purity is obtained.

\subsection{Michel Electron Selection}\label{subsec:Michel_electron_selection}

\begin{figure}[!htpb]
\centering
\includegraphics[width=0.98\linewidth]{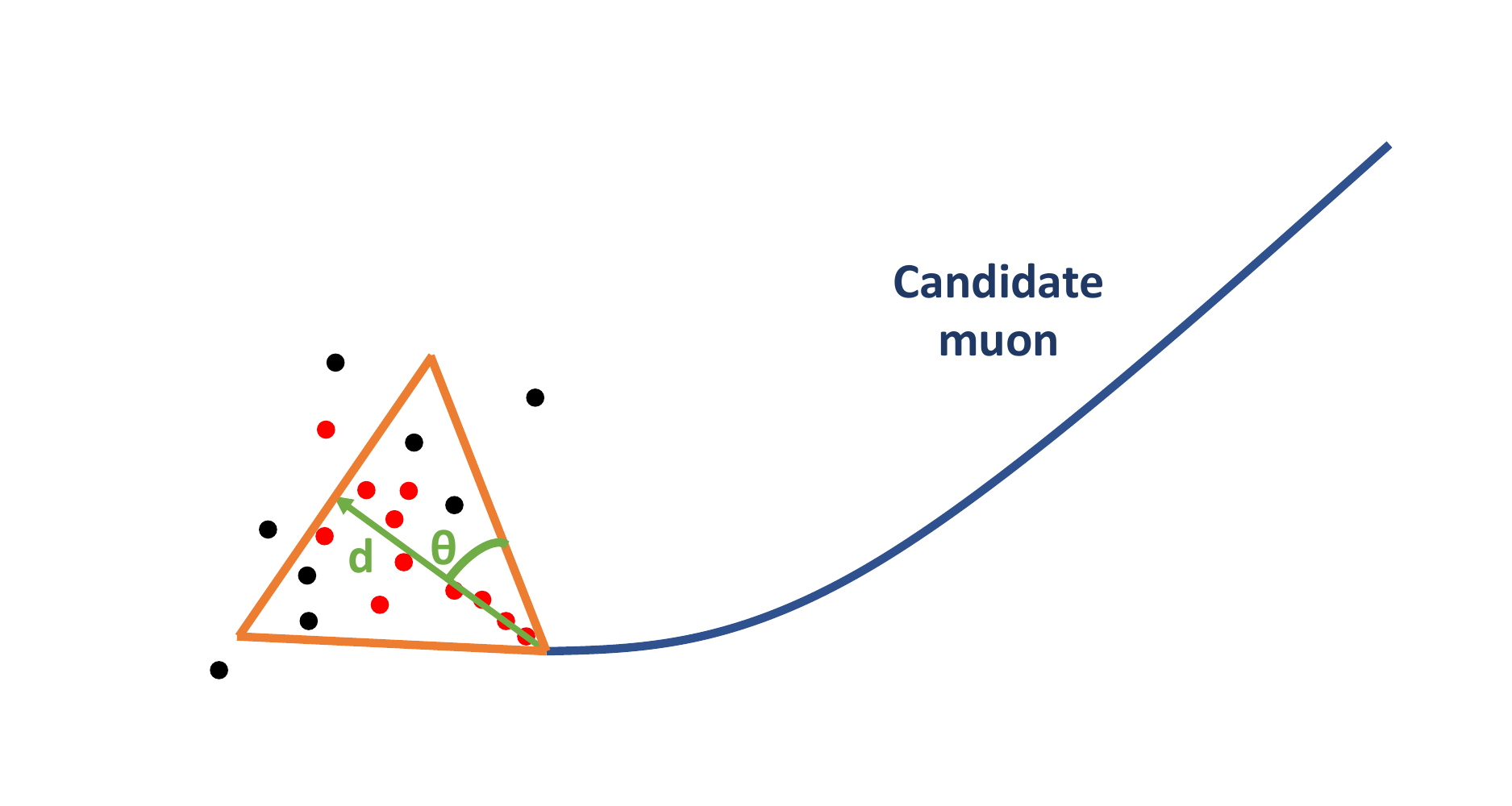}
\caption{Illustration of the cone containment that separates Michel electron hits (red dots) from nearby cosmic ray background events (black dots). Michel electron is defined by the hits starting within 10 cm from the end position of the candidate muon. All the hits contained inside the cone cuts are taken to be the candidate Michel electron hits.}
\label{fig:coneGeo}
\end{figure}

\begin{enumerate}[(i)]

\item The first step in the identification of Michel electrons is to select nearby hits, i.e. hits within 10~cm of the end position of the candidate muon. 
In the collection plane view, these hits must not belong to either the candidate muon or any other track having length $>$10~cm. Nearby hits are counted and events that have between $5$ and $40$ hits around the endpoint of the candidate muon track are considered. These values are optimized to deliver a high sample purity. Furthermore, the reconstructed electron shower around the candidate muon track endpoint is required to start within 10~cm from the candidate muon track end position.
The Michel electron candidate is formed from these selected electron shower hits.  

\item The direction of the candidate Michel electron (obtained by a linear fit to the nearby hits) is compared to the direction of the muon (measured using the last 10~hits in the trajectory). The angle between the directions is required to be less than $130^\circ$ such that events where the candidate Michel electron goes back along the muon are rejected.

\item In the next step, the angle between the collection plane wires and the direction of the candidate Michel electron is calculated. Only those events where the value of this angle is $>10^\circ$ and $<170^\circ$ are selected so that Michel electron candidates that are parallel to the collection plane wires are not included in the data sample. This cut is applied to reject Michel electrons that are parallel to the collection plane wires and therefore may not have well-reconstructed hits.  

\item The final selection criterion, the cone cut, separates Michel electron hits from nearby cosmic rays that may interact in the TPC close to the candidate Michel electron event. A cone around the endpoint of the candidate muon is defined such that any hit that lies within that cone is assumed to belong to the candidate Michel electron. It is required that those hits are not a part of the parent muon or any other track longer than 10 cm. 
A straight line is fit along the nearby collection plane hit distribution (hits within 10 cm distance from the muon endpoint in the collection plane). 
The cone cut is illustrated in Figure \ref{fig:coneGeo}, where the red points represent the Michel electron hits, and the black points are other (non-Michel candidate) hits of the event excluding the hits of the parent muon or any other long ($>$10~cm) tracks.  Using simulation, the cone opening angle $\theta$ is optimized at $70^\circ$ and the cone length $d$ is 20~cm in order to maximize the value of Michel electron hit purity ($83\%$) and hit completeness ($74\%$). The hit purity is defined as the fraction of hits in the reconstructed cone that actually belong to the true Michel electron. The hit completeness is defined as the fraction of true Michel electron hits inside the reconstructed cone.
\end{enumerate}

\subsection{Event Selection Summary}

Table \ref{tab:passingrates} lists the muon passing rates and corresponding statistical uncertainties with respect to the well-reconstructed muon tracks and candidate Michel electrons passing rates with respect to the well-reconstructed muons that satisfy the muon selection criteria for the simulation and data samples. As the focus of this analysis was to select a pure sample of Michel electrons, an estimation of the systematic uncertainties on the passing cosmic ray muon rates was not performed here. The total number of data events in this study that pass all the selection steps is $\sim8300$. 
The total event purity of the selected electron sample from the simulation is found to be $95\%$. The purity here corresponds to the fraction of the true Michel electron events out of all the selected events.
The remaining $5\%$ of events represent different types of background events including those that have a tagged electromagnetic activity from muons (delta-rays or bremsstrahlung photons), in which some random noise hits appear to be reconstructed as candidate Michel electron hits or those in which protons are emitted from argon nuclei because of the muon capture on argon. Isolating background events in the simulation, their energy spectrum is found to be monotonically decreasing, with a low-end cutoff at $\sim$10~MeV. These secondary background events have been characterized elsewhere~\cite{ICARUS_recomb, reconstruction_instruments}.

\begin{table}[!htpb]
\caption{Passing rates for event selection criteria applied to ProtoDUNE simulation and data samples. Quantities with statistical uncertainties for muon selection present the percentage with respect to well-reconstructed muons.  Quantities with statistical uncertainties for Michel electron selection present the percentage with respect to well-reconstructed muons that satisfy the muon selection criteria.}
\label{tab:passingrates}\centering
\renewcommand{\tabcolsep}{1pt}
\begin{tabular}
[c]{l|c|c}
\hline
\textbf{Passing rates} &\textbf{Simulation} & \textbf{Data } \\\hline
Muon selection  & $(27.9\pm0.1)\%$ & $(25.5\pm0.1)\%$ \\
Michel electron selection & $(16.3\pm0.1)\%$ &  $(14.5\pm0.2)\%$ \\
\hline
\end{tabular}
\end{table}

It is important to point out that the DUNE far detector data will be dominated by single $\nu_{\mu}$ or single $\nu_e$ events, where the event selection and reconstruction efficiencies will improve in the absence of nearby cosmic ray background activity, as opposed to the ProtoDUNE-SP case studied in this article. 
It is expected that the muon flux inside ProtoDUNE-SP is on the order of one per $\text{cm}^2/\text{min}$~\cite{Thijs_Miedema_thesis}.
The event selection criteria will be revisited and optimized for the DUNE far detector analyses. An expected muon rate in four modules of DUNE being underground will be about 0.2~Hz with an average muon energy of 283~GeV~\cite{MeV_DUNE_potential}. 

\section{Test and Verification of the Michel Electron Recombination Correction}
\label{sec:recomb_test}

\begin{figure}[!htpb]
\centering
\begin{minipage}{.5\textwidth}
\subfloat{\includegraphics[height=5.0cm,width=1.0\textwidth]{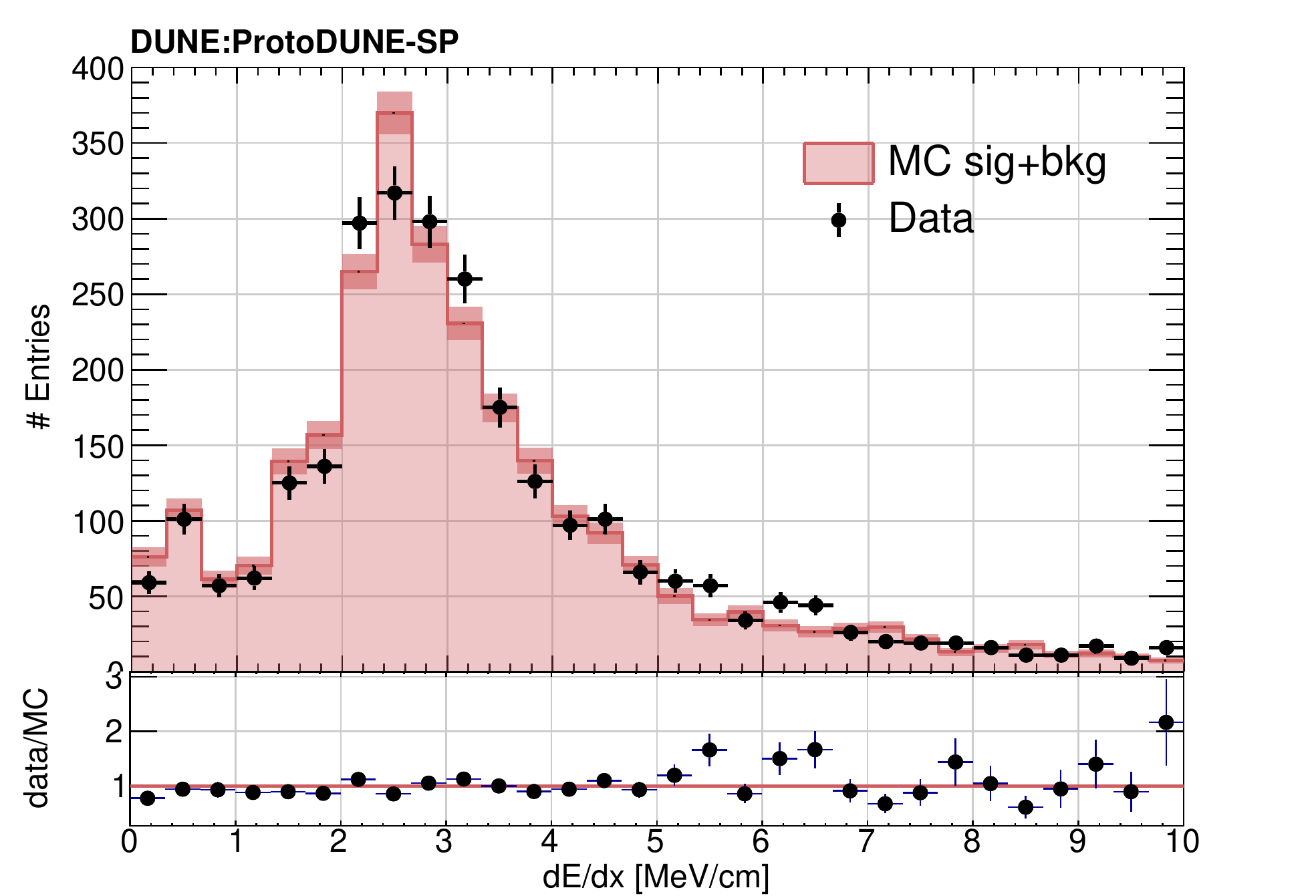}}
\end{minipage}
\begin{minipage}{.5\textwidth}
\subfloat{\includegraphics[height=5.0cm,width=1.0\textwidth]{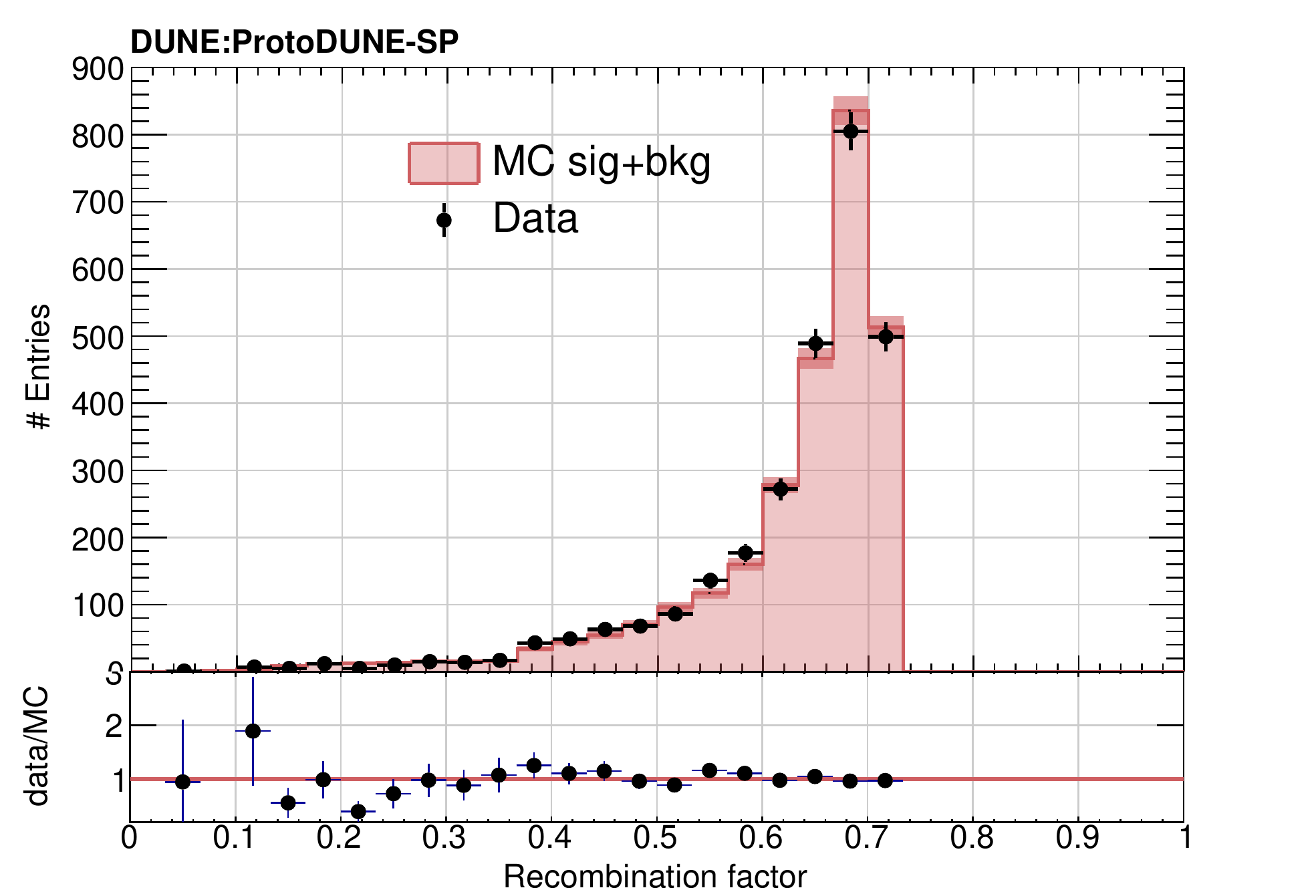}}
\end{minipage}
\caption{Computed Michel electron $dE/dx$ (top); and recombination correction factor (bottom): Data (black points) and MC simulation (red histogram) are compared. }
\label{fig:dedx_r}
\end{figure}

A through-going charged particle will deposit energy in LAr by creating both ionization and excitation. Electron-ion pairs will be produced ($e^-$, Ar$^{+}$), along with excited argon atoms (Ar$^*$). These excited atoms (Ar$^*$) will form excited molecular argon ions, so-called short-lived excimers (Ar$_{2}^*$), through collisions of Ar$*$ with neutral Ar atoms. In addition, the Ar$_{2}^*$ will be also formed by free electrons recombining with surrounding molecular argon ions (Ar$_{2}^+$). These excimers (Ar$_{2}^*$) undergo dissociative decay to their ground state by emitting the vacuum ultraviolet photons known as argon scintillation light~\cite{Michel_lariat, ICARUS_recomb, LArQL}. When the deposited energy is reconstructed using charge alone, as done in the work presented here, only the electrons that escape electron-ion recombination and successfully drift to anode collection wires will be accounted for. $R$ is the recombination factor that describes the fraction of ionization electrons that survive prompt recombination with argon ions before the drift towards the anode plane.
The value of $R$ is critical to energy reconstruction from collected ionization charge, as later described in Equation~\ref{eq:Ecal_muon}.
In this subsection the data-driven recombination correction factor is derived by following the Modified Box model~\cite{box_model}. The Michel electron candidates in this study are selected with the cuts described in Section~\ref{subsec:Michel_electron_selection}.
The electron energy loss per unit length is calculated on an event-by-event basis. 
The value of $dQ/dx$ per event is computed as 
\begin{equation}
  dQ/dx = \frac{Q_{total}}{L},
\end{equation}
where $Q_{total}$ is the total charge deposited determined from the candidate Michel electron hits and $L$ is the 3D displacement from the first to the last hit of the candidate Michel electron. In both data and MC, raw $dQ/dx$ is converted to corrected $dQ/dx$ based on calibration constants derived with the cosmic ray muons~\cite{perf_paper}, as described later by Equation~\ref{eq:Ecal_muon}.
With the Modified Box model~\cite{box_model}, the calibrated $dQ/dx$ value is converted to an average $dE/dx$ for every Michel electron candidate. The average $dE/dx$ distribution of Michel electron candidates is shown in Figure~\ref{fig:dedx_r} (top). The mean value of the $dE/dx$ distribution is 3.25~MeV/cm. 
Finally, the agreement of simulation with data is tested using the recombination correction factor distribution. The recombination factor $R$ is calculated as

\begin{equation}
  \label{eq:recombination_factor}
  R = \frac{ \ln \left( \frac{dE}{dx} \times \beta^{\prime}/\rho E_f + \alpha \right) }{ \frac{dE}{dx} \ \times \beta^{\prime}/\rho E_f },
\end{equation}

where $\alpha$ and $\beta^\prime$ are the Modified Box model parameters which were measured by the ArgoNeuT experiment at an electric field strength of 0.481~kV/cm~\cite{box_model}. The values of $\alpha$ and $\beta^\prime$ are 0.93$\pm$0.02 and 0.212$\pm$0.002~\text{(kV/cm)(g/cm\textsuperscript2)/MeV} respectively. The liquid argon density $\rho$ at a pressure of 124.11~kPa 
is 1.38~\text{g/cm\textsuperscript{3}}, and $E_f$ is the applied electric field. 
Using Equation~\ref{eq:recombination_factor}, $R$ is computed for each event using $dE/dx$ for the event and assuming a constant electric field of 0.5~kV/cm~\cite{electric_field_value}.
The $R$ distribution of Michel electron candidates is shown in Figure~\ref{fig:dedx_r} (bottom). 

The mean value of the recombination factor obtained from the reconstructed data and MC distributions in Figure~\ref{fig:dedx_r} (bottom) is 0.625$\pm$0.020 (stat) and 0.626$\pm$0.020 (stat) respectively. 
Independent of the analysis performed above, the average recombination factor of 0.644 $\pm$ 0.014 (syst) was evaluated based on the ProtoDUNE-SP Geant4 electron simulation~\cite{GEANT}, which incorporates the Modified Box model of the ionization electron recombination and its systematic uncertainty as described in~\cite{box_model}. 
The recombination factor $R$~=~0.644 derived with the simulation comes with a small uncertainty and agrees well with the data-driven value described in this subsection, verifying the simulation-based recombination factor applied in the analysis described in this article. 

\section{Michel Electron Lost Energy Studies}
\label{sec:Eff_studies}

The lost energy is a fraction of energy that is not reconstructed. It corresponds to ionization charge deposits that are missed by either being left below the anode charge readout threshold or left outside the selection cone. This subsection describes the Michel electron lost energy studies performed to quantify the containment of Michel electron events within the applied cone cut and to evaluate the effects of TPC readout thresholds using MC simulation.

\subsection{Michel Electron Hit Completeness} \label{sec:outside_cone}

\begin{figure}[!htpb]
\centering
\includegraphics[height=5.0cm,width=1.0\linewidth]{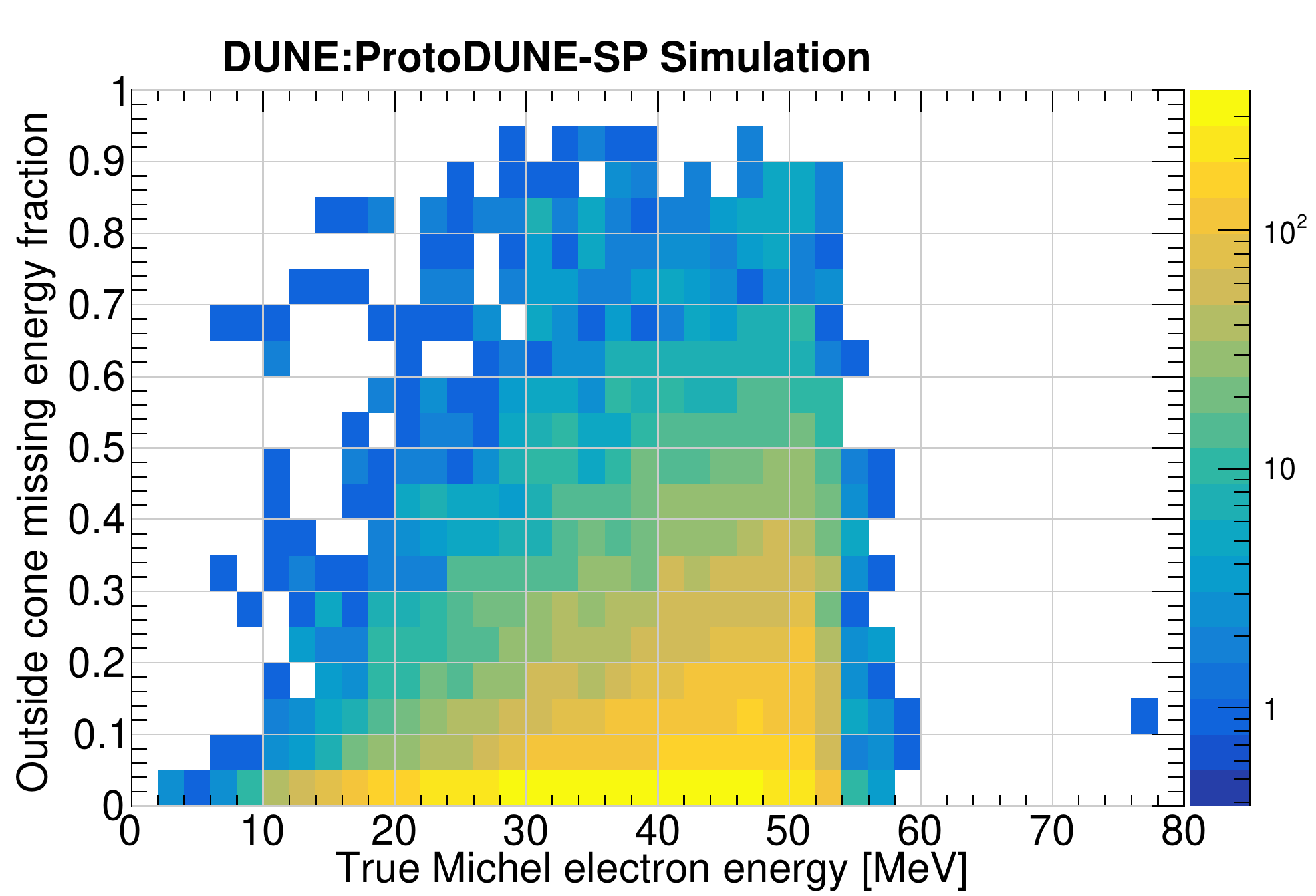}
\caption{True Michel electron energy fraction left outside the selection cone as a function of true Michel electron energy.}
\label{fig:Complteness}
\end{figure}

Figure~\ref{fig:Complteness} shows the fraction of true Michel electron energy left outside the cone as a function of true Michel electron energy per event. 
The average value of the energy loss due to hit incompleteness for the Michel electron sample is 13$\pm$1(stat)\%. It is also evident that the energy loss by hits not captured within the reconstruction cone increases with the Michel electron energy due to the increase of radiative losses. 

\subsection{Michel Electron Hit Reconstruction Threshold}\label{sec:hit_reco_thresh}
\begin{figure}[!htpb]
\centering
\includegraphics[height=5.0cm,width=1.0\linewidth]{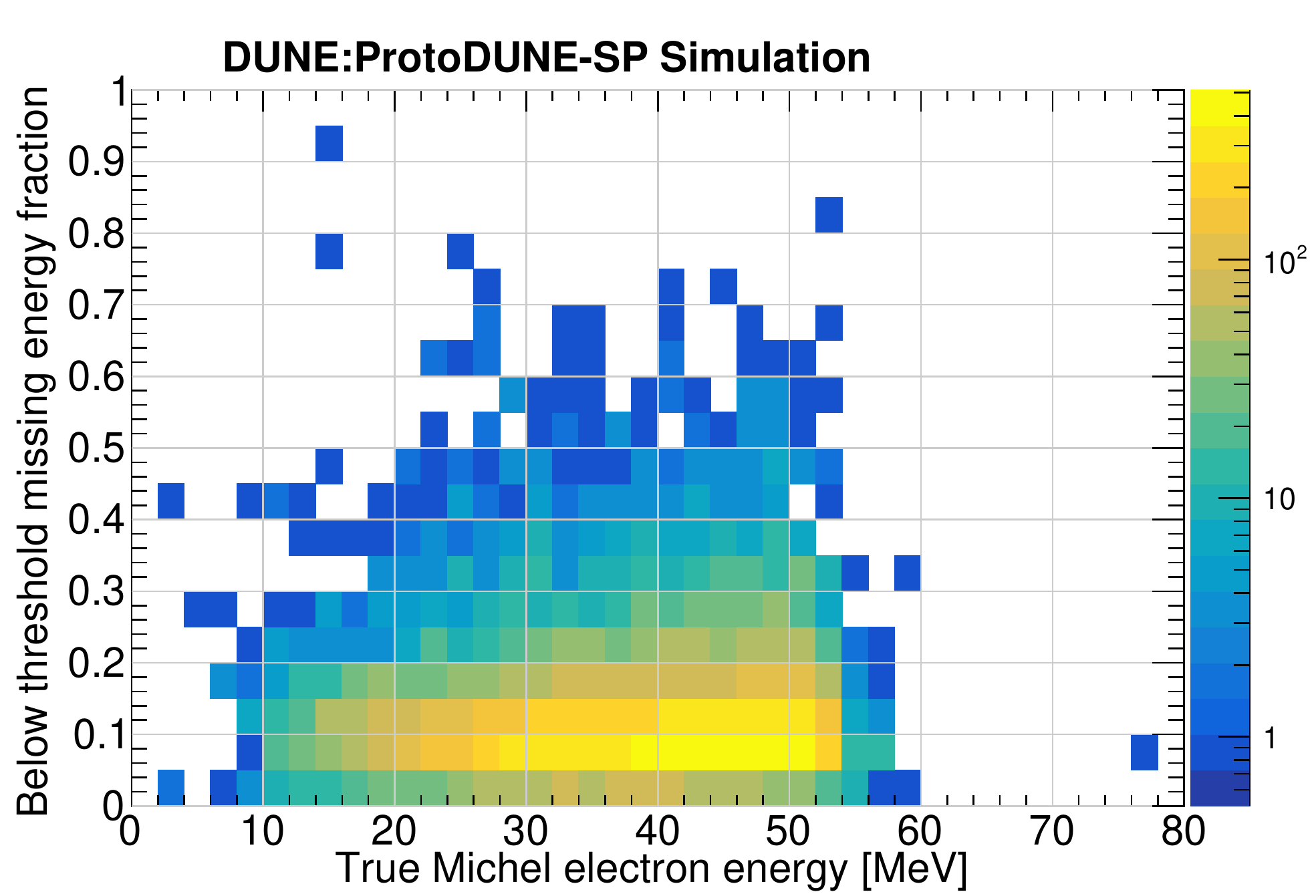}
\caption{True Michel electron energy fraction left below the charge readout threshold as a function of true Michel electron energy.}
\label{fig:missingE}
\end{figure}

In order to avoid random noise from being reconstructed as a particle hit, there is an intrinsic threshold applied to the energy deposited in a given readout channel (wire) per time tick, the value of which is set to $\sim$100~keV/tick. To quantify the impact of the threshold on the Michel electron energy distribution, a study was performed to look at all simulated channels and to estimate the lost energy due to the above-mentioned threshold. 
Figure~\ref{fig:missingE} shows the true Michel electron lost energy fraction as a function of true Michel electron energy per event from this threshold; on average 11$\pm$1(stat)\% of the ionization from Michel electrons is lost due to this threshold. 

Therefore the total of about 24\% of the true energy is not reconstructed from hits outside the cone and below threshold, so only 76$\pm$1(stat)\% of the total energy is captured.

\section{Michel Electron Energy Reconstruction} 

This section describes the procedure of Michel electron energy reconstruction. In the first method (also called as nominal reconstruction) cosmic ray muon data are used to derive calibration constants and corrections~\cite{perf_paper}, which are then applied to reconstructed Michel electron hits. The second approach is based on the well-understood theoretical Michel electron energy spectrum~\cite{original_Michel} where the energy calibration is independent of the muon-based calibration. Finally, the energy resolution effects important for understanding the electron energy in the $5-50$~MeV range in LArTPCs are discussed. 
It should be clarified that the nominal energy reconstruction presented here does not include the lost energy because it cannot be captured in the data with the existing event selection and charge readout threshold. However, potential energy reconstruction improvements with some or all of the lost energy recovered are studied with the simulation to indicate opportunities that might be realized with future DUNE far detector LArTPCs.

\subsection{Muon-based Energy Reconstruction of the Michel Electron Energy Scale}
\label{sec:muon-based-e}

The electron energy, $E$, is calculated from the sum of charges deposited by the corresponding ionization electron hits on the anode plane wires. 
The total reconstructed energy of the Michel electron is given as 

\begin{equation}
\label{eq:Ecal_muon}E = \frac{C_{norm} * W_{ion}}{R*C_{calib}} * \sum_{i = 1}^{N} \Big[\epsilon(X_{i})* \epsilon(Y_{i}, Z_{i})*dQ_{i}\Big]
\end{equation}
where \(dQ_{i}\) (in ADC tick) corresponds to the charge deposited in the $i^\textrm{th}$ hit, and $N$ corresponds to the total number of candidate Michel electron hits.
Note that $dQ/dx$ values along the drift direction are affected by attenuation due to electronegative impurities and by longitudinal diffusion.
Here, $C_{norm}$ is the factor that normalizes the reconstructed $dQ/dx$ values to the average $dQ/dx$ value across anode planes in both drift volumes; \(\epsilon(X_{i})\) represents the drift electron lifetime and the space charge corrections, and \(\epsilon(Y_{i}, Z_{i})\) describes the dead wire correction that is used to remove the non-uniformity in $dQ/dx$ values~\cite{perf_paper}.
In addition, \(W_{ion}\) (= 23.6~eV) is the ionization work function of Argon~\cite{Wion}. A highly pure sample of stopping muons is used in ProtoDUNE to correct for space charge effects and to determine $dQ/dx$~\cite{perf_paper}. From the calibrated $dQ/dx$ values (in~ADC/cm) along the muon track in its MIP region, the $dE/dx$~(in~MeV/cm) values are fitted using the Modified Box model~\cite{box_model} function to correct for the recombination effect with the charge calibration constant $C_{calib}$  as a free parameter in the $\chi^2$  minimization. Therefore $C_{calib}$~(ADC tick/electron) represents the calibration constant that is used to convert the corrected charge deposition (in~ADC) on a hit to energy deposition (in~MeV) on a hit.
It accounts for the electronics gain of the collection-plane wires, the signal processing, as well as detector effects that convert the deposited energy into collected electrons on the wire planes. \(R\)~=~0.644 is the average recombination correction evaluated by the ProtoDUNE-SP GEANT4 simulation based on the Modified Box model~\cite{box_model}, and verified above on an event-by-event basis by selected Michel electron events.
The reconstructed Michel electron energy is evaluated on event-by-event basis using Equation~\ref{eq:Ecal_muon} in which all the calibration corrections are derived from cosmic ray muon data and simulation samples. Therefore, the energy reconstruction applied to the Michel electron sample in this subsection is based on cosmic ray muon calibration.

With the Michel electron energy reconstruction described in Equation~\ref{eq:Ecal_muon}, it is appropriate to evaluate systematic uncertainty contributions to the energy scale. These contributions originate from charge hit ($dQ_i$) association efficiency, the recombination factor ($R$) uncertainty, the theoretical Michel electron versus positron uncertainty, and from the space-charge effects ($\epsilon(X_{i})$) uncertainty.
These uncertainties quantify how well the absolute energy scale of Michel electrons is understood. Systematic uncertainty contributions from $C_{calib}$,  \(\epsilon(Y_{i}, Z_{i})\) and $C_{norm}$ are negligible.
Table~\ref{tab:uncertainties} presents the 
systematic uncertainties on the reconstructed Michel electron energy spectrum. The uncertainties are expressed with respect to the mean energy of the reconstructed Michel electron energy spectrum. Individual contributions are added in quadrature.

\begin{table}[!htpb]
\caption{Michel electron energy spectrum systematic uncertainties estimates from simulation. The uncertainties are expressed with respect to the mean energy of the reconstructed Michel electron energy spectrum.}
\label{tab:uncertainties}\centering
\renewcommand{\tabcolsep}{1pt}
\begin{tabular}
[c]{l|c}
\textbf{Sources of systematic uncertainties}& \textbf{Uncertainty estimates} \\\hline
Hit association efficiency & 4.0\%\\ 
Recombination factor & 2.2\%\\
Michel electron versus positron & 1.7\% \\ 
Space charge effect & 1.4\% \\
\hline
\bf{Total added in quadrature} &  \bf{5.1\%} \\
\end{tabular}
\end{table}

The hit association systematic uncertainty was evaluated by considering the number of candidate Michel electron hits within 10~cm of the muon stopping point in both data and simulation. The difference in the average number of hits in data and simulation was used to vary the MC Michel electron hit distribution. A shift in the mean value of the reconstructed Michel electron energy scale was determined  based on the hit distribution variations. This is the largest  systematic contribution in this analysis with a value of 4.0\%, originating from the requirement to separate electron from muon hits. 
A systematic uncertainty of $2.2\%$ is assigned to the recombination factor resulting from the $2\%$ uncertainty on the Modified Box model~\cite{box_model} parameters and the $1\%$ uncertainty based on the electric field variation from the Modified Box model value of $0.481$~kV/cm to $0.500$~kV/cm in ProtoDUNE. In addition, a test of systematic effects on the use of average constant recombination correction (\(R\)~=~0.644) for all events was tested in our data-driven method described in Section~\ref{sec:recomb_test}. When comparing the reconstructed electron energy using $R$ as an average constant value for all events to the energy using $R$ applied on event-by-event basis, a difference of $<1.0\%$ was found between the two derived energy scales, which was well within the systematic error assigned to the recombination factor $R = 0.644\pm0.014$.
%
\begin{figure}[!htpb]
\centering
\includegraphics[height=6.0cm,width=1.0\linewidth]{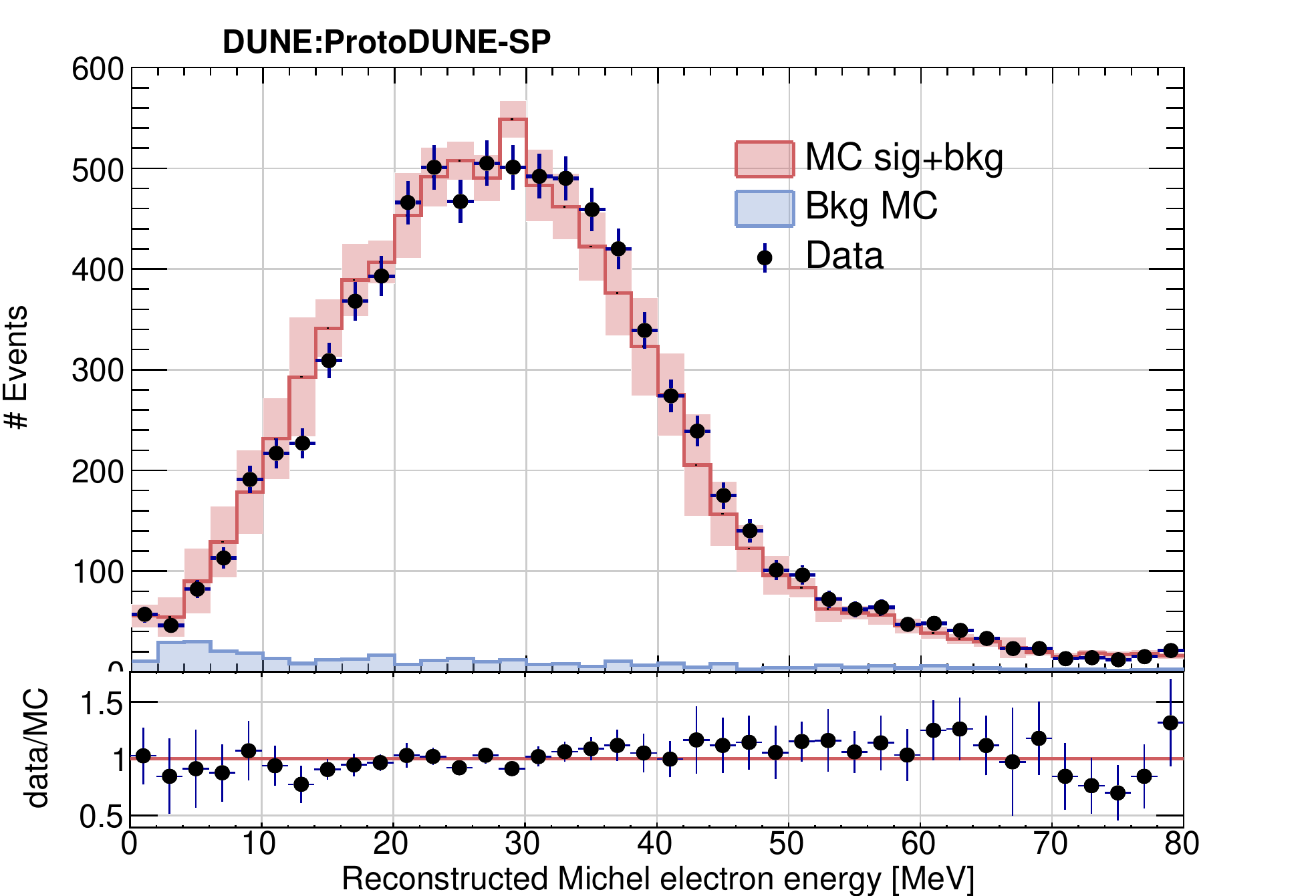}
\centering
\includegraphics[height=6.0cm,width=1.0\linewidth]{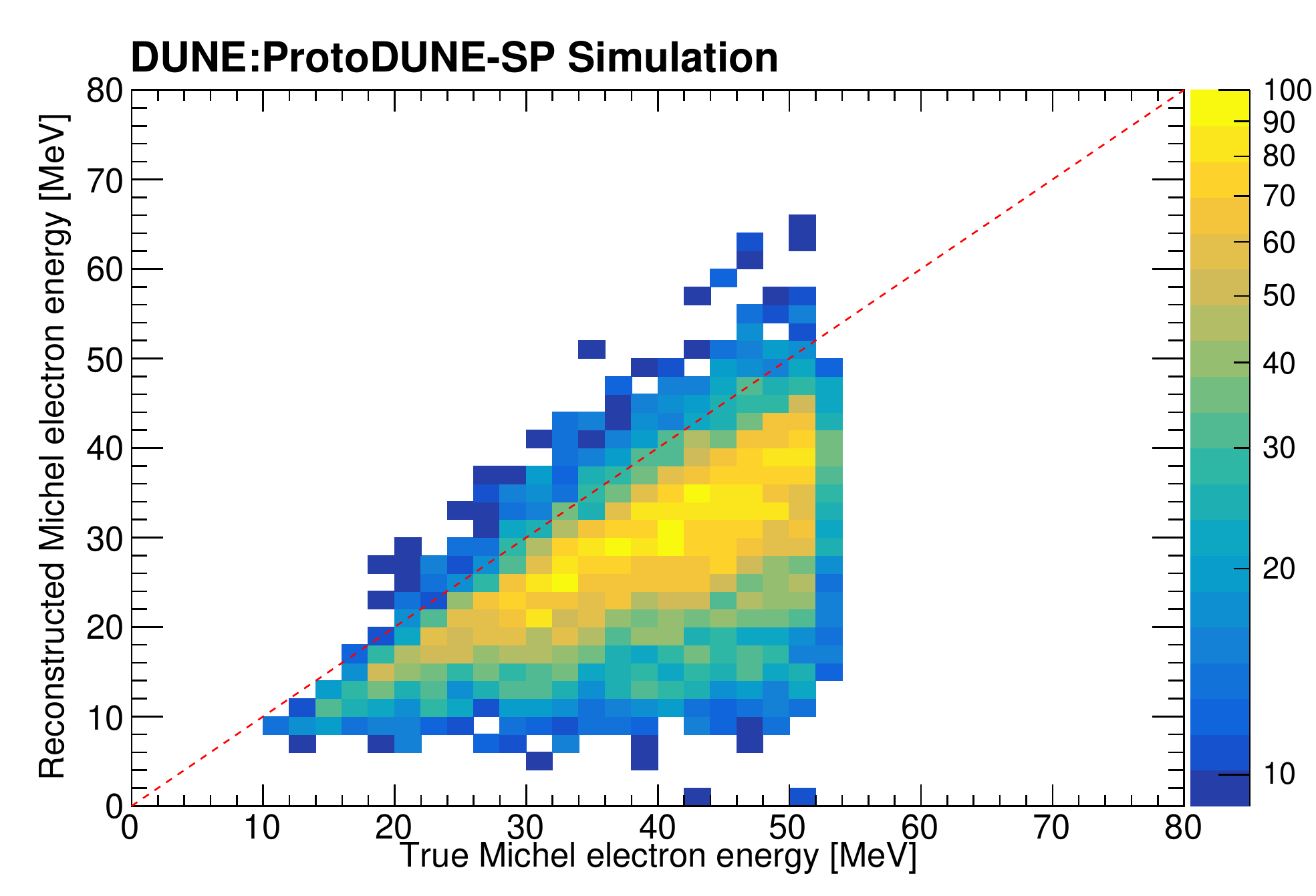}
\caption{Reconstructed Michel electron energy spectrum from ProtoDUNE-SP data and simulation (top); and Michel electron reconstructed energy using stopping muon calibration versus true Michel electron energy (bottom). Bins having a very low number of events are not shown in this plot.}
\label{fig:Michel_energy}
\end{figure}

For the evaluation of the systematic uncertainty from the difference between Michel electrons and positrons, the impact on the true Michel electron energy spectrum is evaluated by considering both electrons' and positrons' energy spectra separately. The systematic uncertainty was evaluated by taking the ratio of the difference between the means of the two distributions of electrons and positrons to the mean of the distribution having both positrons and electrons included. The uncertainty from this systematic contribution is 1.7\%.
The space charge effect is due to the non-uniformity in the electric field due to the low mobility of heavy Ar ions compared to the electrons in the TPC. To quantify the space charge effect systematic uncertainty, simulated data samples with space charge ON and OFF are evaluated. The systematic uncertainty is estimated by taking the percentage difference in the average value of the reconstructed Michel electron energy spectrum from both samples. The space charge affects the mean value of the Michel electron energy distribution by about 1.4\%. 
In conclusion, the total systematic uncertainty on the absolute Michel electron energy scale is estimated to be 5.1\%.

Figure \ref{fig:Michel_energy} (top) presents the reconstructed Michel electron energy spectrum using the muon-based calibration with ProtoDUNE data (in black points), from MC simulation including both signal and background contributions (in red), and from background only MC events (in blue).
The number of MC simulation events is normalized to the number of data events. The data error bars represent the statistical uncertainties. MC simulation error bands include MC statistical and systematic uncertainty contributions added in quadrature. The mean value of the reconstructed Michel electron energy spectra is 29.4$\pm$0.2 (stat)~MeV and 28.7$\pm$0.2 (stat)$\pm$1.4 (syst)~MeV for data and MC simulation, respectively. Relative energy scales of data and MC simulation events agree to within 1.8\%. 
The ratio of data to MC reconstructed energy spectra is flat within statistical and systematic uncertainties. Figure \ref{fig:Michel_energy} (bottom) shows the reconstructed Michel electron energy versus the true Michel electron energy distribution.
These results demonstrate that the Michel electron energy spectrum from data is closely reproduced by the theoretically well-understood Michel electron energy distribution when propagated through the detector simulation and reconstruction.

\subsection{Michel Electron Calibration to True Energy}
\label{sec:true-based-e}
This section describes an alternative approach to estimate the reconstructed Michel electron energy spectrum by using the theoretical Michel electron energy distribution. 
The model applied here assumes a linear relationship between collected charge and reconstructed energy as motivated by the muon-based electron energy reconstruction method described by Equation~\ref{eq:Ecal_muon}.
The charge collected by the collection plane wires is converted to true Michel electron energy by a calibration procedure in which the true Michel electron energy distribution convolved with a resolution function with 
parameters that characterize the electron energy resolution is fit to the charge distribution, using an energy resolution model described later by Equation~\ref{eq:fit_params} and discussed in Section~\ref{sec:resomuonCalib}.
The total reconstructed energy of the Michel electron is given as 
\begin{equation}
\label{eq:Ereco_true}E = \frac{ \sum_{i = 1}^{N} dQ_{i}}{C_{scale}}
\end{equation}
where \(dQ_{i}\) (in ADC tick) corresponds to the charge deposited in the $i^\textrm{th}$ hit, and $N$ corresponds to the total number of candidate Michel electron hits. The calibration scale factor, $C_{scale}$ (ADC tick/MeV), translates the collected charge to reconstructed Michel electron energy.
By using the simulation to relate the true (theoretical) Michel energy spectrum to collected charge, the fit parameters ($C_{scale}=95.2\pm3.1~\text{ADC tick/MeV}$, $p_{0}=0.20\pm0.08$, $p_{1}=2.10\pm0.08$~MeV$^{1/2}$, and $p_{2}=6.85\pm0.29$~MeV are obtained with a Minuit minimization algorithm~\cite{Miniut}. This four-parameter approach matches simulated true and reconstructed energy distributions with the best value of $\chi^2$/ndf~=231/46.
The $C_{scale}$ parameter is then applied on an event-by-event basis to the simulation and data to obtain the updated reconstructed Michel electron energy spectrum. This true energy-based fitting procedure with resolution smearing is used to match the reconstructed charge to true energy, while the energy resolution is characterized in Section~\ref{sec:resomuonCalib}. 

\begin{figure}[!htpb]
\centering
\begin{minipage}{.5\textwidth}
\includegraphics[height=6.0cm,width=1.0\textwidth]{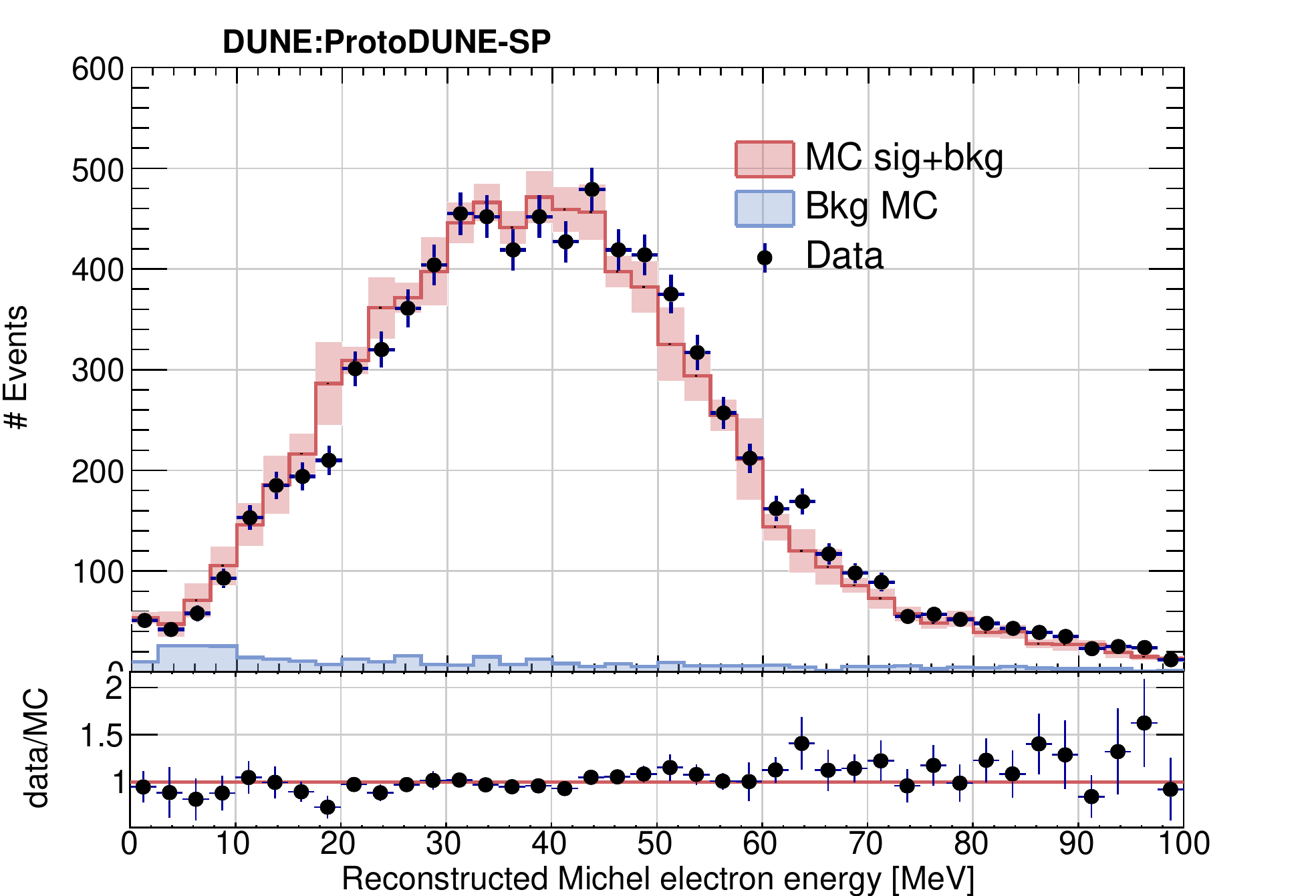}
\end{minipage}
\begin{minipage}{.5\textwidth}
\includegraphics[height=6.0cm,width=1.0\textwidth]{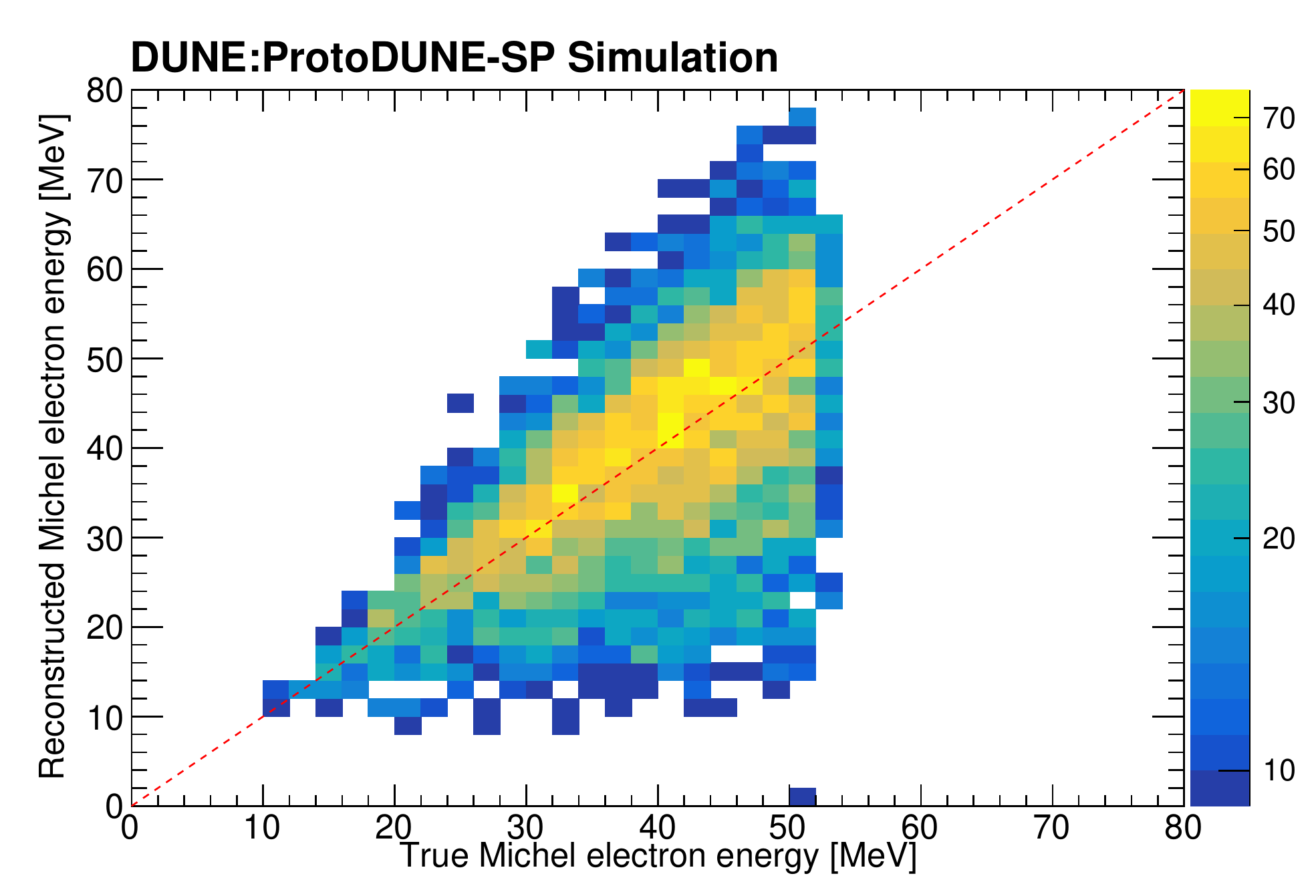}
\end{minipage}
\caption{Michel electron reconstructed energy distribution obtained from the Michel electron charge distribution after dividing by a calibration fitted scale factor for MC simulation (in red) and for data (in black) distribution (top); and Michel electron reconstructed energy using true Michel electron calibration versus the true Michel electron energy (bottom). Bins having a very low number of events are not shown in this plot.}
\label{fig:trueCalib}
\end{figure}

Systematic uncertainties in the truth-based energy scale come from the
need to convert collected nominal charge to energy, and from the impact of Michel electron and positron content in the true energy spectrum.
The corresponding value of the $C_{scale}$ was changed by $\pm1\sigma$ where $\sigma$ denotes the statistical uncertainty associated with its value obtained from the fit. Evaluation of the systematic uncertainty from the spectral difference between Michel electrons and positrons is already described above.
The uncertainties from these systematic contributions turn out to be 2.9\% and 1.7\% respectively. In conclusion, the total systematic uncertainty on the absolute Michel electron energy scale is estimated to be about 3.4\%. 

Figure~\ref{fig:trueCalib} (top) presents the reconstructed Michel electron energy spectrum using the true energy-based calibration with ProtoDUNE data (in black points), from MC simulation including both signal and background contributions (in red), and from the background only MC events (in blue).
The number of MC simulation events is normalized to the number of data events. The data error bars represent the statistical uncertainties. MC simulation error bands include MC statistical and systematic uncertainty contributions added in quadrature. After correcting the reconstructed charge distributions in data and simulation by the same $C_{scale}$ factor, the mean value of the reconstructed Michel electron energy spectra is 40.4$\pm$0.2 (stat)~MeV and 39.1$\pm$0.2 (stat)$\pm$1.3 (syst)~MeV for data and MC simulation, respectively. The energy scales of data and MC simulation events agree to within 2.6\%. The ratio of data to MC reconstructed energy spectra flattens out when the MC energy scale is varied with predicted systematic uncertainties. Figure~\ref{fig:trueCalib} (bottom) shows the reconstructed Michel electron energy versus the true Michel electron energy distribution.

The true energy-based calibration depends on the collected charge only, and is independent of the muon-based calibration constants and of the recombination correction. With the reconstructed charge distribution normalized to directly match the true Michel electron energy distribution smeared with a resolution function, the final reconstructed Michel electron energy distribution does not show the energy offset due to major losses of energy outside the cone and the energy below the hit reconstruction threshold. These losses are incorporated in the value of $C_{scale}$. Therefore the mean of the reconstructed Michel electron energy distribution using the true energy-based calibration method is higher than the one obtained using the muon-based calibration method.

\subsection{Michel Electron Energy Resolution}\label{sec:resomuonCalib}

A proper interpretation of the Michel electron energy resolution requires a complete understanding of the main sources of energy loss. About $24\%$ of the Michel electron energy is lost when using the reconstructed cone-only energy paired with readout threshold effects described in Section~\ref{sec:Eff_studies}.
This is a significant fraction of the energy and a proper understanding of energy underestimation is very important for DUNE and for other LArTPC experiments. It is appropriate to point out that this amount of Michel electron energy loss was previously observed by other LArTPC experiments but up to now, no detailed investigation of the causes for lost energy has been reported~\cite{MicroBooNE_Michel, Michel_lariat}.

\begin{figure}[!htpb]
\centering
\includegraphics[height=6.0cm,width=1.0\linewidth]{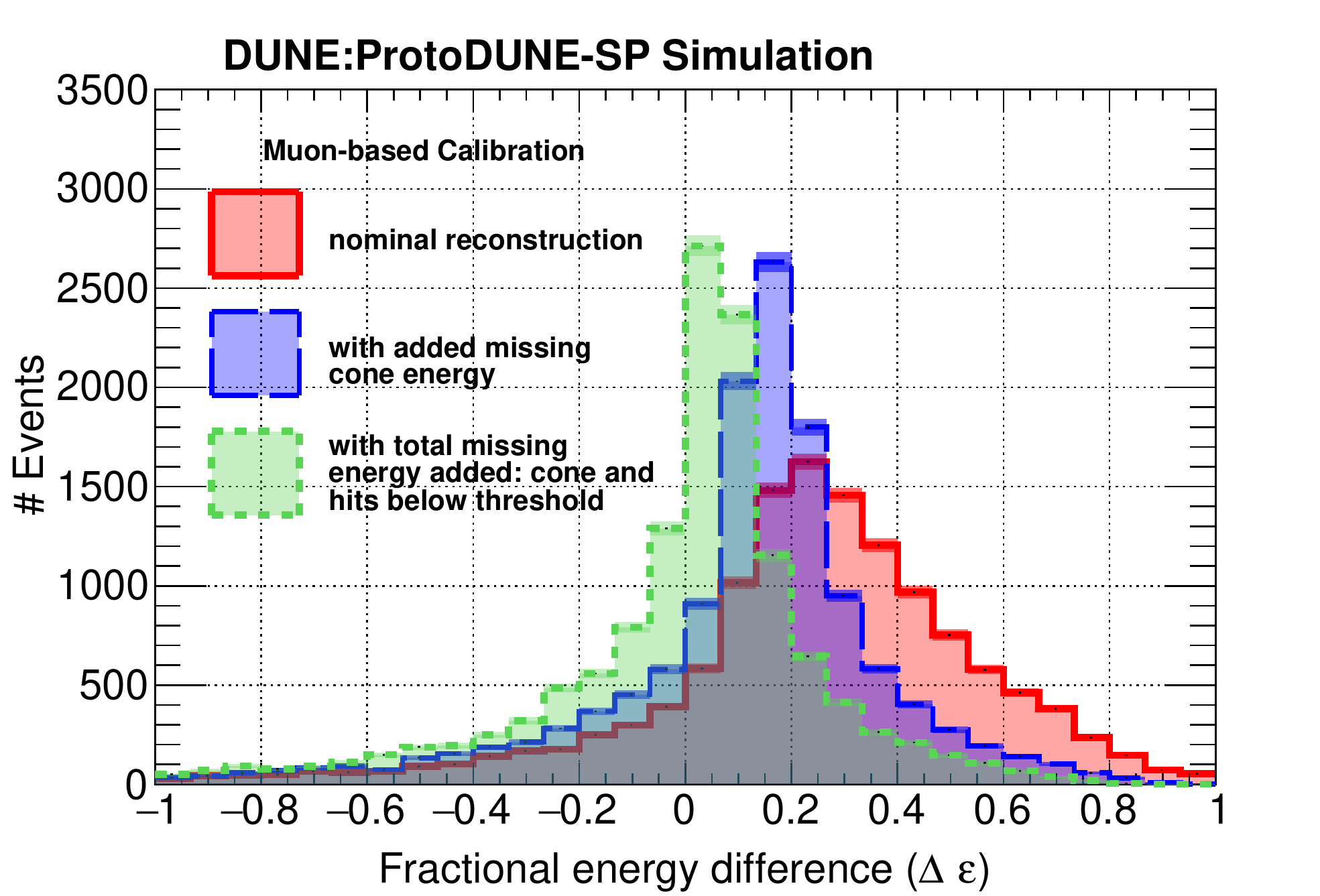}
\caption{$\Delta~\epsilon$ distribution before and after the addition of lost cone and hit reconstruction energies.}
\label{fig:fracenergy}
\end{figure}

The fractional energy difference ($\Delta \epsilon$) per event is defined as
\begin{equation}
    \Delta \epsilon = \frac{(E_{true} - E_{reco})}{E_{true}}
\end{equation}
where $E_{true}$ is the true Michel electron energy and $E_{reco}$ is the reconstructed Michel electron energy per event. 
Figure~\ref{fig:fracenergy} shows various $\Delta\epsilon$ distributions in the simulation: before the addition of any lost energy contribution, called nominal reconstruction (in red), after the addition of the lost energy outside the reconstructed cone (in blue), and after including the additional contribution of lost energy due to hit reconstruction threshold (in green). The $\Delta\epsilon$ peak is closer to zero when both lost energy contributions are added, in contrast to the situation before the addition of lost energy components. 

\begin{figure}[!tpb]
\centering
\begin{minipage}{.5\textwidth}
\includegraphics[height=6.0cm,width=1.0\textwidth]{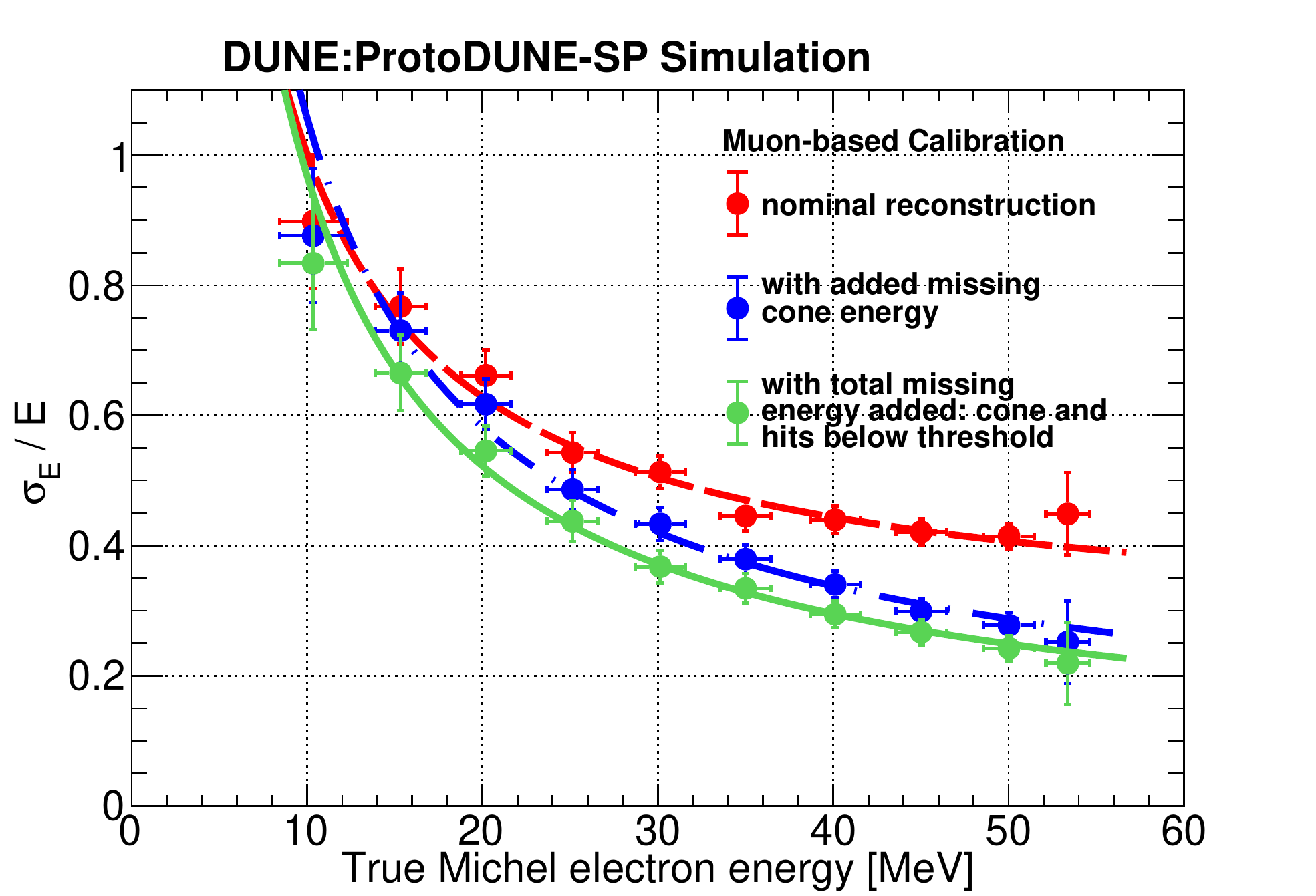}
\end{minipage}
\begin{minipage}{.5\textwidth}
\includegraphics[height=6.0cm,width=1.0\textwidth]{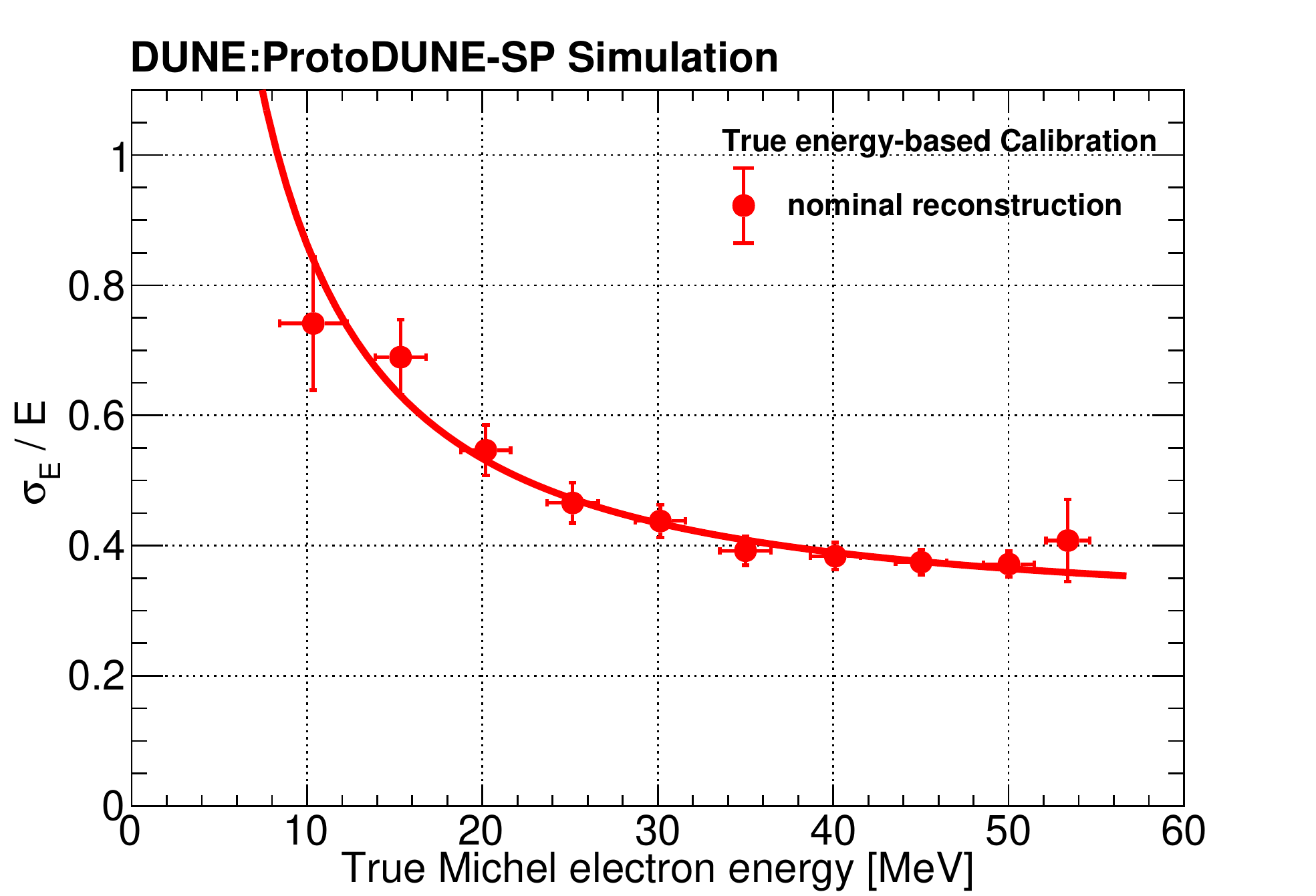}
\end{minipage}
\caption{Michel electron energy resolution as a function of Michel electron true energy using muon-based calibration (top); and the Michel electron energy resolution as a function of Michel electron true energy using true energy-based calibration (bottom).}
\label{fig:energy_resolution}
\end{figure}

For a homogeneous calorimeter such as the ProtoDUNE-SP LArTPC, the energy resolution $\sigma(E)/E$ is expressed by the equation
\begin{equation}
\frac{\sigma(E)}{E} = p_{0} \oplus \frac{p_{1}}{\sqrt{E}} \oplus \frac{p_{2}}{E}
\label{eq:fit_params}
\end{equation}
where $\sigma(E)/E$
is the standard deviation in the reconstructed Michel electron energy divided by the true Michel electron energy distribution in energy bins (shown in Figure~\ref{fig:energy_resolution}) and $E$ is the mean of the reconstructed Michel electron energy distribution obtained in each true energy bin. The terms on the right-hand side are the constant term ($p_0$), the stochastic term ($p_1/\sqrt{E}$) and the noise term ($p_2/E$). The operator \(\oplus\) indicates a sum in quadrature. The constant term describes the resolution losses due to lost energy. The stochastic term incorporates contributions to the energy resolution from the statistical fluctuations in the number of ionization electrons, and scales as 1/$\sqrt{E}$. The noise term accounts for the electronic noise of the collection wires and readout electronics ADCs and scales as $1/E$. 

The Michel electron energy resolution distributions are shown as a function of true Michel electron energy in Figure~\ref{fig:energy_resolution}. The x-axis points represent the mean values of the Michel electron true energy bins and the horizontal error bars correspond to the standard deviation of each true Michel electron energy bin; the statistical uncertainty on the fit to $\sigma(E)/E$ values (from Equation~\ref{eq:fit_params}) is shown along the vertical axis. Table~\ref{tab:energy_resolution} presents the mean values of the Michel electron energy for various samples as well as the values obtained for the parameters in Equation~\ref{eq:fit_params}, as obtained from a least-squares fit. 

\begin{table*}[!htpb]
\caption{Average Michel electron energy for reconstructed data and MC spectra, and fit parameter values obtained for the Michel electron energy resolution model (Equation~\ref{eq:fit_params}) from MC.}
\label{tab:energy_resolution}
\centering
\renewcommand{\tabcolsep}{1pt}
\begin{tabular}
[c]{l||c|c||c|c|c}
\textbf{Energy reconstruction} & \multicolumn{2}{c||}{ \textbf{Mean energy [MeV]}} & \multicolumn{3}{c}{\textbf{Energy resolution parameters}}\\\
& \textbf{Data} & \textbf{MC} & \textbf{Constant ($p_0$)} & \textbf{Stochastic ($p_1$ [$\sqrt{\text{MeV}}$])} & \textbf{Noise ($p_2$ [$\text{MeV}$])}\\\hline
\begin{tabular}{l}
    Muon-based \\
    (nominal reconstruction)     
  \end{tabular} &
  \begin{tabular}{l}
    $29.4\pm0.2 (\text{stat})$ 
  \end{tabular} &
  \begin{tabular}{l}
    $28.7\pm0.1 (\text{stat})\pm1.4 (\text{syst})$
  \end{tabular} &
  \begin{tabular}{l}
    $0.26\pm0.12$
  \end{tabular} &
  \begin{tabular}{l}
    $1.91\pm0.93$
  \end{tabular} &
  \begin{tabular}{l}
    $7.54\pm3.05$
  \end{tabular}
  \\
  \hline
\begin{tabular}{l}
    Muon-based \\
    (total lost energy added)     
  \end{tabular} &
  \begin{tabular}{l}
    ...
  \end{tabular} &
  \begin{tabular}{l}
    $38.3\pm0.1 (\text{stat})\pm1.9 (\text{syst})$
  \end{tabular} &
  \begin{tabular}{l}
    $0.00\pm0.15$
  \end{tabular} &
  \begin{tabular}{l}
    $1.24\pm1.22$
  \end{tabular} &
  \begin{tabular}{l}
    $8.86\pm0.94$
  \end{tabular}
  \\
  \hline
\begin{tabular}{l}
    True energy-based \\
    (nominal reconstruction)     
  \end{tabular} &
  \begin{tabular}{l}
    $40.4\pm0.2 (\text{stat})$ 
  \end{tabular} &
  \begin{tabular}{l}
    $39.1\pm0.1 (\text{stat})\pm1.3 (\text{syst})$
  \end{tabular} &
  \begin{tabular}{l}
    $0.29\pm0.09$
  \end{tabular} &
  \begin{tabular}{l}
    $1.21\pm1.28$
  \end{tabular} &
  \begin{tabular}{l}
    $7.17\pm2.80$
  \end{tabular}
  \\
  \hline
\end{tabular}
\end{table*}

Constant and stochastic terms are affected by the resolution losses due to the lost energy. 
Assuming the lost energy contributions described in the MC study (Section~\ref{sec:Eff_studies}) are added back to the energy balance, the resolution improves, as quantified in Table~\ref{tab:energy_resolution}. 
Figure~\ref{fig:energy_resolution} shows the energy resolution fit (Equation~\ref{eq:fit_params}) applied to both muon-based and true energy-based calibrated Michel electron energy with the results consistent within the statistical uncertainties. The energy resolution of $\sim40\%$ is derived at 50~MeV when using the nominal (i.e. without lost energy recovered from the MC) muon-based calibration method. If the lost energy contributions from outside the cone and energy below hit reconstruction threshold are added, the resolution improves relative to the nominal case: 
the constant term becomes very small and the resolution becomes limited  by the stochastic term from about 50~MeV energy.
For parameters in Table~\ref{tab:energy_resolution} the correlation coefficients are consistent between muon-based and true energy-based calibration methods with the constant term $p_{0}$ being highly anti-correlated with the stochastic term $p_{1}$. The following observations can be  made:
\begin{enumerate}[(i)]
\item In the nominal cosmic muon-based calibration, the collected (visible) charge is corrected by calibration constants and converted to reconstructed energy. Since the mean of the true Michel electron energy spectrum is at 38.4~MeV, the nominal energy reconstruction recovers $\sim$75\% of the total deposited energy.

\item The energy resolution constant term ($\sim$26\%) describes the resolution losses due to the lost energy.
In the MC simulation, it is possible to 
add the lost energy to the reconstructed energy balance. 
In this case, the energy resolution drops from about $40\%$ to $25\%$ at $50$~MeV. The lost energy recovery is not possible in the collected data set with the existing charge readout threshold and applied selection (cone cut) optimized to reduce backgrounds.

\item If the lost energy is accounted for, as performed in the simulation, the stochastic term decreases 
from 1.91 to about 1.24. 
The stochastic term described here may suggest that the energy resolution at a few percent level might be achievable for the DUNE far detector in the few GeV electron neutrino energy range, assuming negligible lost energy and noise contributions to the resolution. These potential improvements would have important implications for low-energy electrons expected for supernova neutrinos, and also for the few GeV scale electrons to be observed in the DUNE far detector from $\nu_{\mu} \rightarrow \nu_e$ oscillation.

\item In the 10-50~MeV energy range
relevant to solar or supernovae measurements, 
the noise term currently dominates. In order to improve the sensitivity of these measurements, one needs to improve understanding of the noise effects. 


\item By construction, the true energy-based Michel electron reconstruction is in good agreement with the theoretical Michel electron energy as presented in Figure~\ref{fig:trueCalib}. As a consequence, the mean values of the Michel electron energy distribution in data (40.4~MeV) and in MC simulation (39.1~MeV) are in close agreement with the theoretical Michel electron energy spectrum with the mean at 38.4~MeV. The method does not recover the energy resolution loss but accounts for the average lost energy.

\end{enumerate}

\section{Summary}

This article describes the event selection and energy reconstruction of low-energy electrons in the DUNE prototype ProtoDUNE-SP LArTPC. A high-purity (95\%) Michel electron event sample is selected and used to calibrate the electron energy scale, and to quantify the effects of the detector response to low-energy electrons including readout electronics threshold effects. 
The event selection techniques for cosmic ray muons and Michel electrons have been developed. 
The selected Michel electron sample was used to verify the recombination correction factor and the data and MC simulation agreement was presented based on the Modified Box recombination model. 

Two complementary energy reconstruction techniques to calibrate the Michel electron energy spectrum are described. The ``cosmic-muon'' based calibration is based on a model-dependent recombination correction and relies on the calibration constant derived from cosmic ray muon measurements. On the other hand, the ``true-energy'' based calibration method is based on the theoretical Michel energy spectrum and is independent of any correction applied in the muon-based energy reconstruction. 
An excellent agreement between data and simulation for the Michel electron energy spectrum to within 2\% and 3\% using muon-based calibration and the true energy-based Michel electron calibration respectively have been observed. 
Once a sample of the low-energy electrons is collected in the far detector, it will be calibrated to its true energy by converting reconstructed visible energy based on the relationship presented in Figure~\ref{fig:Michel_energy} (bottom) for the muon-based method and Figure~\ref{fig:trueCalib} (bottom) for the true-based calibration method.  The two methods will cross-check each other and may quantify a lost energy.
As part of this analysis, the estimates for systematic uncertainties on the Michel electron energy spectrum are presented. 
The dominant contribution to the systematic uncertainty comes from the difference in the hit association effects close to the candidate muon end position in the data and simulation.

This article also reports the sources of the lost energy and quantifies the effect of each of those sources separately. The lost energy coming from charge readout threshold effects and from the event selection is evaluated to be about 11\% and 13\% respectively. 
The energy resolution is quantified in this analysis.
In the nominal “cosmic-muon” based calibration, the collected charge is corrected by calibration constants and converted to reconstructed energy. While the mean of the true Michel electron energy spectra is at 38.4~MeV, the nominal energy reconstruction mean is at 28.7~MeV due to the lost energy effect. 
After the addition of the lost energy in MC, the constant term approaches zero and the stochastic resolution term improves by 35\%.
In such a case the energy resolution improves from
about 40\% to 25\%, at 50~MeV.
Assuming the lost energy is fully recovered with negligible noise contributions, the DUNE far detector may reach the energy resolution defined by stochastic term alone (Equation~\ref{eq:fit_params}) consistent with low-energy physics goals of DUNE~\cite{dune-physics-TDR} and other LArTPCs~\cite{Snowmass_paper}.
These results demonstrate the capabilities of the ProtoDUNE-SP (and ultimately the DUNE far detector) to detect and reconstruct electrons with energies up to $\sim{50}$~MeV. 

For further improvements it will be important to understand if the charge collection in DUNE far detector can operate at lower thresholds and noise levels to further improve the energy resolution.
The DUNE far detector data will be dominated by single muon and electron events, where the electron event selection and reconstruction efficiencies will improve in the absence of nearby cosmic ray background activity. As a result the event selection will be further optimized for the DUNE far detector analyses. 
Understanding of energy resolution and its potential improvements will have important implications for electrons from $\nu_e$ interactions 
in the DUNE far detector. Analysis of low-energy neutrino interactions in DUNE will benefit from a combination of muon-based energy calibration, Michel electron true energy-based calibration, and detailed MC modeling to characterize the energy resolution with potential energy losses.

\section{Acknowledgment}
The ProtoDUNE-SP detector was constructed and operated on the CERN Neutrino Platform. We thank the CERN management for providing the infrastructure for this experiment and gratefully acknowledge the support of the CERN Experimental physics (EP), Beams (BE), Technology (TE), Engineering (EN), Information Technology (IT), and Industry, Procurement and Knowledge Transfer (IPT) Departments for NP04/ProtoDUNE-SP. This document was prepared by the DUNE collaboration using the resources of the Fermi National Accelerator Laboratory (Fermilab), a U.S. Department of Energy, Office of Science, HEP User Facility. Fermilab is managed by Fermi Research Alliance, LLC (FRA), acting under Contract No. DE-AC02-07CH11359. This work was supported by CNPq, FAPERJ, FAPEG and FAPESP, Brazil; Foundation for Innovation (CFI), Institute of Particle Physics (IPP) and NSERC, Canada; CERN; MŠMT, Czech Republic; ERDF, H2020-EU and MSCA, European Union; CNRS/IN2P3 and CEA, France; INFN, Italy; FCT, Portugal; NRF, South Korea; Comunidad de Madrid (CAM), Fundación “La Caixa” and MICINN, Spain; SERI and SNSF, Switzerland; TÜBİTAK, Turkey; The Royal Society and UKRI/STFC, United Kingdom; DOE and NSF, USA. This research used resources of the National Energy Research Scientific Computing Center (NERSC), a U.S. Department of Energy Office of Science User Facility operated under Contract No. DE-AC02-05CH11231.


\newpage
\begin{thebibliography}{99}                                 

\bibitem{cp-violation-davidson} S.~Davidson, E.~Nardi and Y.~Nir, Leptogenesis, \href{https://www.sciencedirect.com/science/article/pii/S0370157308001889?via\%3Dihub}{\color{blue} Phys. Rep. {\bf 466}, 105 (2008)}.
  
\bibitem{cp-violation-fukugita} M.~Fukugita and T.~Yanagida, Baryogenesis Without Grand Unification, \href{https://www.sciencedirect.com/science/article/pii/0370269386911263?via\%3Dihub}{\color{blue} Phys. Lett. B {\bf 174}, 45 (1986)}. 



\bibitem{dune_osc_paper} B.~Abi {\it et al.} (DUNE Collaboration), Long-baseline neutrino oscillation physics potential of the DUNE experiment, \href{https://link.springer.com/article/10.1140/epjc/s10052-020-08456-z}{\color{blue} Eur. Phys. J. C {\bf 80}, 978 (2020)}.


\bibitem{dune-physics-TDR} B.~Abi {\it et al.} {DUNE Collaboration}, Deep Underground Neutrino Experiment (DUNE), Far Detector Technical Design Report, Volume II: DUNE Physics, \href{https://arxiv.org/abs/2002.03005}{\color{blue} arXiv:2002.03005}.

\bibitem{dune-intro-TDR} B.~Abi {\it et al.} (DUNE Collaboration), Deep Underground Neutrino Experiment (DUNE), Far Detector Technical Design Report, Volume I Introduction to DUNE, \href{https://iopscience.iop.org/article/10.1088/1748-0221/15/08/T08008}{\color{blue} J. Instrum. {\bf 15}, no. 08, T08008 (2020)}.

\bibitem{dune-nd-cdr} A.~Abed~Abud {\it et al.} (DUNE Collaboration), Deep Underground Neutrino Experiment (DUNE) Near Detector Conceptual Design Report, \href{https://www.mdpi.com/2410-390X/5/4/31}{\color{blue} Instruments 2021, {\bf 5}, 31}.

\bibitem{mass_hierarchy} X.~Qian and P.~Vogel, Neutrino mass hierarchy, \href{https://www.sciencedirect.com/science/article/pii/S0146641015000307?via\%3Dihub}{\color{blue} Prog. Part. Nucl. Phys. {\bf 83}, 1 (2015)}.

\bibitem{dune_bsm} B.~Abi {\it et al.} (DUNE Collaboration), Prospects for beyond the Standard Model physics searches at the Deep Underground Neutrino Experiment, \href{https://link.springer.com/article/10.1140/epjc/s10052-021-09007-w}{\color{blue} Eur. Phys. J. C {\bf 81}, 322 (2021)}.
   
\bibitem{dune_sn} B.~Abi {\it et al.} (DUNE Collaboration), Supernova Neutrino Burst Detection with the Deep Underground Neutrino Experiment, \href{https://link.springer.com/article/10.1140/epjc/s10052-021-09166-w}{\color{blue} Eur. Phys. J. C {\bf 81}, 423 (2021)}.

\bibitem{dune_solar} F.~Capozzi, S.~W.~Li, G.~Zhu and J.~F.~Beacom, DUNE as the Next-Generation Solar Neutrino Experiment, \href{https://journals.aps.org/prl/abstract/10.1103/PhysRevLett.123.131803}{\color{blue} Phys. Rev. Lett.  {\bf 123}, 131803 (2019)}.

\bibitem{protodune-paper} A.~Abed~Abud {\it et al.} (DUNE Collaboration), Design, construction and operation of the ProtoDUNE-SP Liquid Argon TPC, \href{https://iopscience.iop.org/article/10.1088/1748-0221/17/01/P01005}{\color{blue} J. Instrum. {\bf 17}, P01005 (2022)}.
  	
\bibitem{original_Michel} L.~Michel, Interaction between Four Half-Spin Particles and the Decay of the $\mu$-Meson, \href{https://iopscience.iop.org/article/10.1088/0370-1298/63/5/311}{\color{blue} Proc. Phys. Soc. London Sect. A {\bf 63} 514 (1950)}.

\bibitem{MicroBooNE_Michel} R.~Acciarri {\it et al.} (MicroBooNE Collaboration), Michel electron reconstruction using cosmic-ray data
from the MicroBooNE LArTPC, \href{https://iopscience.iop.org/article/10.1088/1748-0221/12/09/P09014}{\color{blue} J. Instrum. {\bf 12}, P09014 (2017)}.

\bibitem{Michel_lariat} 
W.~Foreman {\it et al.} (LArIAT Collaboration), Calorimetry for low-energy electrons using charge and light in liquid argon, \href{https://journals.aps.org/prd/abstract/10.1103/PhysRevD.101.012010}{\color{blue} Phys. Rev. D {\bf 101}, 012010 (2020)}.

\bibitem{Michel_ArgoNeuT} 
R.~Acciarri {\it et al.} (ArgoNeuT Collaboration), Demonstration of MeV-scale physics in liquid argon time projection chambers using ArgoNeuT, \href{https://journals.aps.org/prd/abstract/10.1103/PhysRevD.99.012002}{\color{blue} Phys. Rev. D {\bf 99}, 012002 (2019)}.

\bibitem{MicroBooNE_stream}
P.~Abratenko \textit{et al.} (MicroBooNE Collaboration), The continuous readout stream of the MicroBooNE liquid argon time projection chamber for detection of supernova burst neutrinos, \href{https://iopscience.iop.org/article/10.1088/1748-0221/16/02/P02008}{\color{blue} J. Instrum. \textbf{16}, P02008 (2021)}.

\bibitem{Snowmass_paper} D.~Caratelli {\it et al.}, Low-Energy Physics in Neutrino LArTPCs, Contribution to Snowmass, \href{https://arxiv.org/abs/2203.00740}{arXiv:2203.00740}. 
  
\bibitem{dune-fd-sp-TDR} B.~Abi {\it et al.} (DUNE Collaboration), Deep Underground Neutrino Experiment (DUNE), far detector technical design report, Volume IV: Far detector single-phase technology, \href{https://iopscience.iop.org/article/10.1088/1748-0221/15/08/T08010}{\color{blue} J. Instrum. {\bf 15}, T08010 (2020)}. 

\bibitem{electric_field_value} S.~Kubota, M.~Hishida, M.~Suzuki, and J.-z.~Ruan~(Gen), Dynamical behavior of free electrons in the recombination process in liquid argon, krypton, and xenon, \href{https://journals.aps.org/prb/abstract/10.1103/PhysRevB.20.3486}{\color{blue} Phys. Rev. B {\bf 20}, 3486 (1979)}.
  
\bibitem{arapuca} A.~A.~Machado and E.~Segreto, ARAPUCA a new device for liquid argon scintillation light detection, \href{https://iopscience.iop.org/article/10.1088/1748-0221/11/02/C02004}{\color{blue} J. Instrum. {\bf 11}, C02004 (2016)}.

    
\bibitem{cern-nu-platform} F.~Pietropaolo, Review of Liquid-Argon Detectors Development at the CERN Neutrino Platform, \href{https://iopscience.iop.org/article/10.1088/1742-6596/888/1/012038}{\color{blue} J. Phys. Conf. Ser.  {\bf 888}, 012038 (2017)}. 

\bibitem{perf_paper} B.~Abi {\it et al.} (DUNE Collaboration), First results on ProtoDUNE-SP liquid argon time projection chamber performance from a beam test at the CERN Neutrino Platform, \href{https://iopscience.iop.org/article/10.1088/1748-0221/15/12/P12004}{\color{blue} J. Instrum. {\bf 15}, P12004 (2020)}.

\bibitem{protodune_CNN}
A.~Abed~Abud \textit{et al.} (DUNE Collaboration), Separation of track- and shower-like energy deposits in ProtoDUNE-SP using a convolutional neural network, \href{https://link.springer.com/article/10.1140/epjc/s10052-022-10791-2}{\color{blue} Eur. Phys. J. C {\bf 82}, 903 (2022)}.


\bibitem{attenuation_length} J.~Calvo \textit{et al.}, Measurement of the attenuation length of argon scintillation light in the ArDM LAr TPC, \href{https://www.sciencedirect.com/science/article/pii/S0927650516302341?via\%3Dihub}{\color{blue} Astropart. Phys. {\bf 97}, 186 (2018)}.

\bibitem{Corsika} D.~Heck, J.~Knapp, J.~N.~Capdevielle, G.~Schatz, and T.~Thouw, CORSIKA: A Monte Carlo code to simulate extensive air showers, Report No. FKZA-6019, 1998, \href{https://
inspirehep.net/literature/469835}{\color{blue} https://
inspirehep.net/literature/469835}.

\bibitem{GEANT} S.~Agostinelli {\it et al.} (GEANT4 Collaboration), GEANT4--a Simulation Toolkit,
\href{https://www.sciencedirect.com/science/article/pii/S0168900203013688?via\%3Dihub}{\color{blue} Nucl. Instrum. Methods  Phys. Res., Sect. A {\bf 506}, 250 (2003)}. 

\bibitem{LArsoft} E.~D.~Church, LArSoft: A Software Package for Liquid Argon Time Projection Drift Chambers, \href{https://arxiv.org/abs/1311.6774}{\color{blue} arxiv:1311.6774}.

\bibitem{GEANT_webpage} Production threshold expressed in range, {\color{blue} \url{https://geant4.web.cern.ch/sites/default/files/geant4/collaboration/working_groups/electromagnetic/gallery/cutinrange/summary.html}}.

\bibitem{pand_package} J. ~S. ~Marshall and M. ~A. ~Thomson, The Pandora Software Development Kit for Pattern Recognition, \href{https://link.springer.com/article/10.1140/epjc/s10052-015-3659-3}{\color{blue} Eur. Phys. J. C {\bf 75}, 439 (2015)}.

\bibitem{reconstruction_instruments} M. ~Szydagis {\it et al.}, A Review of Basic Energy Reconstruction Techniques in Liquid Xenon and Argon Detectors for Dark Matter
and Neutrino Physics Using NEST, \href{https://www.mdpi.com/2410-390X/5/1/13}{\color{blue} Instruments {\bf 5}, 13 (2001)}.

\bibitem{NOvA_seasonal} M.~A.~Acero {\it et al.} (NOvA Collaboration), Seasonal variation of multiple-muon cosmic ray air showers observed in the NOvA detector on the surface, \href{https://journals.aps.org/prd/abstract/10.1103/PhysRevD.104.012014}{\color{blue} Phys. Rev. D {\bf 104}, 012014 (2021)}.

\bibitem{ICARUS_recomb} S.~Amoruso {\it et al.} (ICARUS Collaboration), Study of electron recombination in liquid argon with the ICARUS TPC, \href{https://www.sciencedirect.com/science/article/pii/S0168900204000506?via\%3Dihub}{\color{blue} Nucl. Instrum. Methods {\bf 523}, 275 (2004)}.

\bibitem{Thijs_Miedema_thesis} Thijs Miedema, Stopping Muons at ProtoDUNE, Radboud University Nijmegen, Netherlands, Master Thesis, 2020,
{color{blue} \url{https://www.ru.nl/highenergyphysics/theses/master-theses/}}.

\bibitem{MeV_DUNE_potential} G.~Zhu, S.~W.~Li, and J.~F.~Beacom, Developing the MeV potential of DUNE: Detailed considerations of muon-induced spallation and other backgrounds, \href{https://journals.aps.org/prc/abstract/10.1103/PhysRevC.99.055810}{\color{blue} Phys. Rev. C {\bf 99}, 055810 (2019)}. 

\bibitem{LArQL} F.~Marinho,  L.~Paulucci, D.~Totani, and F.~Cavanna, LArQL: A phenomenological model for treating light and charge generation in liquid argon, \href{https://iopscience.iop.org/article/10.1088/1748-0221/17/07/C07009}{\color{blue} J. Instrum. {\bf 17} C07009 (2022)}.


\bibitem{box_model} R.~Acciarri {\it et al.} (ArgoNeuT Collaboration), A Study of Electron Recombination using Highly Ionizing Particles in the ArgoNeuT Liquid Argon TPC, \href{https://iopscience.iop.org/article/10.1088/1748-0221/8/08/P08005}{\color{blue} J. Instrum. {\bf 8}, P08005 (2013)}.

\bibitem{Wion} E.~Shibamura, A.~Hitachi, T.~Doke, T.~Takahashi, S.~Kubota, and M.~Miyajima, Drift velocities of electrons, saturation characteristics of ionization and W-values for conversion electrons in liquid argon, liquid argon-gas mixtures and liquid xenon, \href{https://www.sciencedirect.com/science/article/pii/0029554X75903274?via\%3Dihub}{\color{blue} Nucl. Instrum. Methods {\bf 131}, 249 (1975)}.

\bibitem{Miniut} F.~James and M.~Roose, Minuit - a system for function minimization and analysis of the parameter errors and correlations, \href{https://www.sciencedirect.com/science/article/pii/0010465575900399?via\%3Dihub}{\color{blue} Comput. Phys. Commun. {\bf 10} 343 (1975)}.


\end{thebibliography}
\end{document}